\documentclass[a4,11pt,amssymbols]{article}
\usepackage{tikz}
\usepackage{subcaption}
\usepackage{amsthm}
\usepackage{array}
\usepackage{amsmath}
\usepackage[colon]{natbib} 
\usepackage{blindtext}
\usepackage{lscape}
\usepackage{subcaption}
\usepackage{eucal}[mathcal]
\usepackage{longtable}
\usepackage{lipsum} 
\usepackage[toc,page]{appendix}
\usepackage{caption}
\usepackage{subcaption}
\usepackage{graphicx}
\usepackage{hyperref}
\usepackage{xurl}
\hypersetup{citecolor=blue,linkcolor=blue,urlcolor=red}
\usepackage{float}
\usepackage{tikz,pgfplots}
\hypersetup{
    colorlinks=true,
    linkcolor={blue},
    filecolor=blue,
    urlcolor=blue,
    citecolor=blue,
    pdftitle={Sharelatex Example},
    }

\usepackage[colon]{natbib}
\hypersetup{citecolor=blue,linkcolor=blue,urlcolor=blue}

\newtheorem{defn}{Definition}

\newtheorem{pro}{Proposition}

\usepackage[hmargin=1in,vmargin=1in]{geometry}

\usepackage[hmargin=1in,vmargin=1in]{geometry}

\begin{document}
\title{Uniform-Price Auctions with Pre-announced Revenue Targets: Evidence from China's SEOs}

\author{Shenghao Gao\thanks{School of Accounting, Southwestern University of Finance and Economics, Sichuan, China. Corresponding author: email gaosh@swufe.edu.cn.}  \hspace{0.5cm} Peyman Khezr\thanks{ Department of Economics, School of Economics, Finance and Marketing, Royal Melbourne Institute of Technology (RMIT), Victoria, Australia.}
 \hspace{0.5cm} Armin Pourkhanali\thanks{Department of Finance, School of Economics, Finance and Marketing, Royal Melbourne Institute of Technology (RMIT), Victoria, Australia.} }
\date{}

\maketitle
\begin{abstract}
This study explores the performance of auctions in China's seasoned equity offering (SEO) market, both theoretically and empirically. In these auctions, issuers must commit to a pre-announced revenue target and a maximum number of shares available for auction. We use a common value framework to analyze this auction mechanism, detailing its operation, share allocation, and pricing. The theoretical findings suggest that when buyers bid truthfully, the seller's optimal strategy is to set the total share quantity equal to the target revenue divided by the reserve price. We demonstrate that committing to a target revenue results in a higher level of truthful bidding compared to a standard uniform-price auction without any revenue commitment. We empirically test our theoretical findings using data from China's SEO markets. First, we assess the impact of various issuer strategies on firm-level SEO discounts, categorizing scenarios based on share availability and target revenue. We find that the scenario where the reserve price times the share quantity matches the target revenue is the most optimal for sellers. Second, we examine bidding behavior and auction performance, showing that China's SEO uniform-price auction performs exceptionally well. Specifically, the actual issue prices are only 0.029 below the truthful case prices, indicating that the revenue raised is still close to what would have been achieved with truthful bids.

\vspace{1ex} \noindent \textit{Keywords}: SEO; uniform-price; auction; Institutional investors. \\[1ex]

\end{abstract}

\section{Introduction}

Auctions play a critical role in financial markets, providing a transparent and efficient mechanism for price discovery and capital allocation. They are used in various contexts, such as government bond issuances, initial public offerings (IPOs), and seasoned equity offerings (SEOs). While several studies investigate the performance of auctions in the primary bond market, especially government bond issuance markets, the literature on auction performance in the primary equity market remains scarce. This study aims to fill this gap by investigating the impact of revenue targets on SEO auctions in China.

SEOs are particularly significant as they allow companies that have already gone public to issue additional shares and raise capital. This process enables firms to access the wealth of the capital markets to fund expansion, reduce debt, or finance new projects. China's SEO market is among the largest in the world, with an approximate annual value of \$93.1 billion. The China Securities Regulatory Commission (CSRC) mandates certain SEOs to determine the offering price and allocate shares through a uniform price auction mechanism.\footnote{During our sample period from 2007 to 2019, there were 4,636 SEOs. Of these, 136 were rights offerings, accounting for 2. 93\%, and 102 were public offerings, accounting for 2.20\%. The majority, 4,398, were private placements, representing 94.87\% of the total. Within these private placements, 1,883 were issued using a uniform-price auction mechanism, comprising 42.81\%.} While the uniform-price auction is commonly used for bond issuance and other financial securities \citep{hortaccsu2021empirical}, its application in new share sales remains infrequent \citep{jagannathan2015share}. Several countries that initially adopted this method in primary equity markets eventually stopped using it \citep{jagannathan2015share}. There is thus a noticeable lack of empirical research on the efficacy of uniform-price auctions in the primary equity market, largely attributed to the insufficiency of observable samples for study. China’s SEO market presents a unique opportunity to assess the effectiveness of these auctions. Issuers in this market are required to disclose detailed bidder information during the auction process, providing granular data on investor bidding behavior. This transparency allows for a thorough evaluation of the effectiveness of uniform-price auctions. 

The auction mechanism in China's SEO market is a uniform-price auction  with a predefined target revenue. Under this mechanism, issuers initially establish a target revenue, set a reserve price, and determine a maximum number of shares for auction. The auction then commences to achieve the target revenue within the constraints of the reserve price and share limit. If the target revenue is attainable with the sales of shares at a price above the reserve and below the share limit, the mechanism allocates only the necessary number of shares at the market-clearing price. If not, the auction ends at the set limits, undershooting the target revenue. Sellers under this mechanism can strategically determine the total shares available so that, when multiplied by the reserve price, the resulting value is either below, equal to, or above the target revenue. 

Despite the significance of China's SEO market, very little is known about the variations applied to the auction and the bidding behavior by buyers given those variations such as target revenue. In this study, we apply a common value theoretical framework to analyze China's SEO auctions, detailing the mechanism's operation, share allocation, and pricing rule. Our theoretical analysis suggests that when buyers bid truthfully, the seller's optimal strategy is to set the total share quantity such that its product with the reserve price equals the target revenue. We further demonstrate that the mechanism adopted in China's SEO market diminishes the incentives of bidders for untruthful bidding and fosters a higher level of truthful bidding compared to the standard uniform-price auction with a fixed supply. 

We tested these theoretical predictions using a machine learning technique and a multiple regression model. First, we apply machine learning techniques as our identification strategy, which is important for establishing a clear cause-and-effect relationship between the variables. This approach enables us to isolate the effects of the variables of interest on SEO discounts while minimizing the influence of confounding factors. In the next step of our analysis, we investigate the impact of various bidding auction policies on firm-level SEO discounts. 
Based on the choice of the total shares available and the amount of money to raise, we have either of the three possibilities: (1) available shares exceed the target revenue, signalling a willingness to sell at the reserve price (Up group); (2) available shares are insufficient to meet the target revenue, indicating confidence that the auction price will exceed the reserve price (Down group); (3) available shares match the target revenue, showing contentment with selling all or some shares based on the auction clearing price (Middle group). We found that the middle group, where the reserve price times the share quantity closely aligns with the target revenue, is most advantageous for sellers. This result is consistent with our theoretical predictions of the seller’s optimal strategy. In a further step of the analysis, we empirically investigate the bidding behavior and the performance of the auction mechanism in China's SEO market. Given that our current data include bids and auction outcomes for every auction, we first construct a hypothetical case to forecast truthful bids (aggregate demand). In particular, we construct the truthful aggregate demand for each auction based on these first bids.\footnote{This analysis relies on existing results in auction theory which indicate bidding equal to the value for the first unit is a weakly dominant strategy in a uniform-price auction \citep{Krishna2009}.} We use this benchmark to investigate the auction price if the bidders were to submit truthful bids. Then, we use this benchmark to construct the truthful equilibrium outcome with the total number of shares available, creating a metric for a scenario where a standard uniform-price auction without the revenue target variation is conducted. Finally, we analyse the performance of the auction, based on two different truthful bidding (hypothetical and real scenarios). We find that China's SEO variation of the uniform-price auction performed reasonably well in terms of revenue generation.  The price difference ratio between the hypothetical and the actual issue price was 0.029, and the actual revenue was only around 3\% below the optimal scenario, showing that the revenue raised closely matched what would have been achieved with truthful bids.
This indicates that the revenue generated in these auctions is nearly equivalent to an ideal truthful bid scenario.

The contribution of this paper is threefold. First, we contribute to the theoretical literature that examines the performance of auctions in securities issuance by formally investigating a uniform-price auction with a target revenue. Although uniform price auctions have been investigated before \citep{Back2001, Damianov2010, McAdams2007, Khezr2020}, we are not aware of any previous study investigating the uniform price auction with a commitment to the preannounced revenue target.\footnote{For a comprehensive survey of prior studies see \cite{Khezr2021}.} Previous studies examine the effects of bidder multi-unit demand (\cite{back1993auctions}; \cite{wang2002auctioning}; \cite{ausubel2014demand}), bid discreteness (\cite{kremer2004underpricing}; \cite{Kastl2011}), dealer information sharing (\cite{nyborg1996discriminatory}, or information asymmetry between primary dealers and other investors (\cite{hortaccsu2018bid}) on auction results. This study contributes to this series of articles in the literature by examining the impact of pre-announced target revenue on the performance of uniform price auctions in the primary equity market. Our evidence shows that the mechanism adopted in China's SEO market diminishes the incentives for untruthful bidding and fosters a higher level of truthful bidding compared to the standard uniform-price auction with fixed supply. 

Second, we empirically investigate the performance of SEO auctions in China where uniform price auctions are widely used in equity markets. Previous studies examine the performance of uniform price auctions by comparing them with other issuance mechanisms or using a structural estimation methodology (e.g. \cite{nyborg1996discriminatory}; \cite{goldreich2007underpricing}; \cite{Kastl2011}; \cite{hortaccsu2018bid}). However, the existing studies predominantly focus on the performance of a standard uniform price auction compared to other auctions such as the pay-as-bid auction for the primary bond markets. There is a lack of empirical studies on the performance of non conventional formats of the uniform price auction such as the one in China's SEO. Using unique data from China's SEO market, this study contributes to the literature by investigating the optimal strategies of sellers and the overall performance of uniform price auctions when the seller sets a target revenue.

Third, we contribute to existing empirical literature by providing evidence that uniform price auctions perform well in primary equity markets, in sharp contrast to the theoretical wisdom of previous studies. Although uniform price auctions are frequently used to issue bonds and other financial securities \citep{hortaccsu2021empirical}, their use in the sale of new shares is relatively rare. Various nations that initially implemented this auction method in their primary equity markets have subsequently abandoned it \citep{jagannathan2015share}. Theoretical studies attribute the limited use of uniform price auctions for new share sales to the complexity of the auction, which poses significant challenges even to experienced bidders trying to mitigate the winner's curse \citep[e.g.][]{sherman2005global, jagannathan2015share}. This complexity is exacerbated by the inherent difficulty in valuing new shares and the fluctuating number of bidders across different auctions \citep{sherman2005global,jagannathan2015share}. 



The paper is structured as follows. Section \ref{SecBack} reviews the relevant literature and provides background information. Section \ref{SecTheory} develops a theoretical model and proposes two hypotheses. Section \ref{SecData} analyzes SEO data using machine learning techniques for preliminary analysis and visualization. Section \ref{SecEmp} employs regression methods to test the hypotheses and conduct additional statistical analysis and further explores bidder behavior in more detail, examining the strategies employed within the auctions. Finally, Section \ref{ConSec} concludes the paper with key findings and remarks.

\section{Institutional background and related literature}\label{SecBack}

\subsection{Institutional background}
SEO is a method for public firms to raise additional capital by issuing new shares. Rights offerings, public offerings, and private placements are three typical types of SEOs. Rights offerings allow companies to raise capital by offering new shares to current shareholders. Public offerings involve selling new shares to all potential investors. Private placements enable public firms to obtain additional equity capital by offering new shares to specific investors. There are two types of private placements in China: competitive bidding and fixed price. While some literature equates China's private placements with U.S. Private Investment in Public Equity (PIPE) (e.g., \cite{lin2020wealth}), there are distinctions.  Fixed-price private placements resemble U.S. PIPE as they target pre-determined entities and set prices through negotiations. However, competitive bidding private placements diverge from U.S. PIPE as they utilize a uniform price auction to finalize issuance prices and investors, enabling all potential investors to bid.

On May 8, 2006, the China Securities Regulatory Commission (CSRC) issued the \textit{Securities Issuance Management Measures for Listed Companies}, officially allowing public firms to refinance equity through private placements. On September 19, 2006, the \textit{Securities Issuance and Underwriting Management Measures} enacted by the CSRC came into effect, stipulating that public firms could choose either a uniform-price auction bidding method or a fixed-price method to determine the issuance price during private placements. On September 17, 2007, the CSRC released the \textit{Detailed Implementation Rules for the Non-Public Issuance of Shares by Listed Companies}, further clarifying the implementation details of bidding, pricing, and allocation in private placements. According to these rules, if the targets of a private placement are determined before submission for CSRC approval or exchange registration, the listed company should negotiate the issuance price with the targets before announcing the board's pre-scheme. If the targets are not determined before submission, the issuance price must be determined through a uniform price auction mechanism.

When adopting the auction mechanism for private placements, listed companies must specify in their prospectuses submitted to the CSRC or the exchange a reserve price, a target revenue, and a cap on the number of shares to be issued. Before February 14, 2020, the reserve price was 90\% of the average price of the 20 trading days before the pricing benchmark date, which could be the date of the board's pre-scheme announcement, the date of the shareholder meeting announcement, or the first day of the issuance period. After February 14, 2020, the reserve price was adjusted to 80\% of the average price of the 20 trading days before the pricing benchmark date, set as the first day of the issuance period. After obtaining CSRC approval, the listed company and the lead underwriter, within the validity period of the approval (6 months before February 14, 2020, and 12 months afterwards), select the issuance timing to provide subscription invitations for bidding to qualified specific targets. The recipients of the subscription invitation should include investors who have submitted letters of intent after the board announcement, the company's top 20 shareholders, at least 20 mutual funds, at least 10 securities companies, and at least 5 insurance institutions. After receiving investor bid prices and quantities, the listed company and the lead underwriter establish a demand curve and set the market-clearing price, where supply equals demand, as the final issuance price. All investors with bids not lower than the issuance price qualify for allocation. Before February 14, 2020, the final number of allottees could not exceed 10. Afterwards, this was adjusted to no more than 35. Before February 14, 2020, investors who obtained shares were not allowed to transfer their subscribed shares for 12 months post-issuance. After this date, the lock-up period for allotted investors was adjusted to 6 months.

To illustrate, Figure 1 plots the issuance process for a private placement using the auction mechanism: decision by the board, vote by the shareholder, approval by the watchdog, bidding by the investor, and pricing by the issuer.  Due to intense government interference in China, private placements 
\begin{figure}[h]
\centering
\includegraphics[width=0.95\textwidth, height=7.5cm]{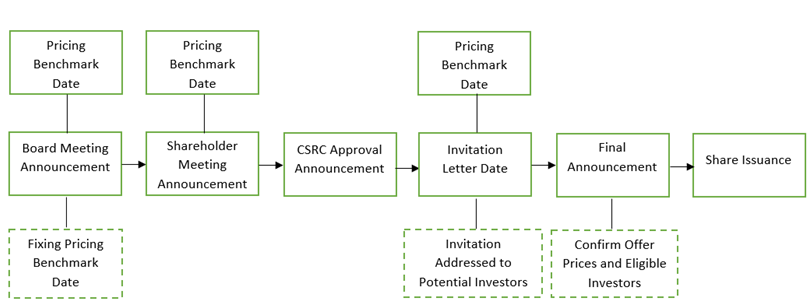}
\caption{Timeline of private equity placements.} 
\label{figSEO}
\end{figure}
manifest several unique institutional characteristics. First, CSRC permission has to be granted before a private placement, whereas public firms are exempt from such a regulatory requirement in other countries (e.g., the US and Singapore). Second, bid/offer prices in an auction cannot be smaller than 90\% or 80\% of the average share price over the 20 trading days before the pricing benchmark date. In principle, the board meeting date, the shareholder meeting date, and the invitation letter date are eligible to act as the pricing benchmark date. Third, the number of successful subscribers for each private placement is capped at 10 or 35, although the issuer has to engage with the top 20 existing blockholders, more than 20 mutual funds, more than 10 brokerages, and more than 5 insurance firms. Thus, this arrangement lays a groundwork for triggering the competition among bidders, conforming to our prior reasoning. Fourth, private placements are a protracted process lasting at least one year until completion (Fonseka et al., 2014). Fifth, the lock-up period covers 12 or 6 (36 or 18) months for external (internal) investors, which is much longer compared to that in developed markets.

\subsection{Literature Review}
The issuance of securities necessitates a pricing mechanism to identify initial investors, determine their share allocations, and establish the pricing. The three predominant approaches are auctions, book building, and fixed-price public offerings \citep{jagannathan2015share}. In the auction mechanism, issuers, typically with the support of underwriters, gather data on the desired price and quantity from investors via competitive bidding. They then set the issue price and distribute the securities based on established rules \citep{mcafee1987auctions}. With book building, the underwriter and issuer initially collect demand information from investors, then independently decide the issue price and allocations based on the collected data. For the fixed-price method, the issue price is negotiated by the underwriter and issuer, followed by investor subscriptions at this price. In the case of over-subscription, securities are distributed to investors either on a proportional basis or through a lottery.

The two most commonly used auctions with regards to the payment rule are the discriminatory price and the uniform-price auctions.\footnote{Note that in theory there are several possibilities for payment rules in auctions. However, discriminatory price and uniform-price are the only ones used in practice in financial markets \citep{Khezr2021}. } In a discriminatory price auction, investors who bid above the market clearing price are allocated shares and must pay according to their respective bid prices.  In a uniform price auction, while investors still need to bid above the market clearing price to be allocated shares, every successful bidder pays the same market clearing price, regardless of their initial bid. Uniform price auctions can further be categorized into two types based on their pricing methodologies: standard and adjusted uniform price auctions. In a standard uniform price auction, underwriters and issuers establish the market clearing price utilizing the bidding data from investors gathered during the inquiry phase, which then serves as the final issuance price. Shares are allocated to all investors whose final bids meet or exceed this market clearing price. However, in the adjusted uniform price auction, which are sometimes called “dirty Dutch” auctions, underwriters and issuers have the flexibility to set the issuance price below the market clearing price, using the investor bid information from the inquiry stage. Here, investors who place final bids above the market clearing price receive share allocations, which are distributed either proportionally or via a lottery system (\cite{degeorge2010auctioned}; \cite{gao2020differences}; and \cite{qian2024initial}).

A significant body of theoretical literature, tracing back to the foundational works of \cite{friedman1960program} and \cite{vickrey1961counterspeculation}, explores the design of pricing mechanisms in the primary bond market. Friedman advocates for the uniform price auction mechanism in the Treasury market, highlighting its advantages: all winning bidders pay an identical price, which diminishes the winner's curse effect, fosters increased investor participation, curtails the propensity for bid shading, boosts issuance efficiency, and reduces financing costs. \cite{vickrey1961counterspeculation} also supports the uniform price auction mechanism but stipulates that the advantages of uniform price auctions predominantly apply to situations where each bidder is interested in no more than one unit and where collusion is absent among the participants.

Building on these insights, a series of research studies has investigated the optimal selection of mechanisms for government issued securities, focusing on maximizing seller revenue and optimizing resource allocation \citep{hortaccsu2018bid}. \cite{back1993auctions} demonstrate that the uniform-price auction is susceptible to a phenomenon called \textit{demand reduction}, where bidders lower their bids for subsequent units to minimize the clearing price, potentially down to the reserve price. Since then, it has become well-known that demand reduction is one of the major flaws of the uniform-price auction \citep{Krishna2009}.

Auctions for securities are typically modeled as common value auctions. In these models, bidders receive signals about the actual value of the items \citep{milgrom1982theory}. The concept of the winner's curse arises in common value auctions, where the winner tends to overestimate the value of the item based on their received signal. \cite{ausubel2014demand} extend this concept to multi-unit auctions, introducing the \textit{generalized winner's curse}. Consequently, the generalized winner's curse provides an additional incentive for bidders to reduce their bids in common value uniform-price auctions.

Several studies have attempted to modify the design of uniform-price auctions to mitigate demand reduction. For instance, \cite{Back2001} show that changing the supply rule can almost eliminate demand reduction. Specifically, if the auctioneer commits only to a maximum quantity of supply and determines the exact amount of supply after bidders submit their bids, demand reduction could be penalized by lower supply and potentially higher prices. In their mechanism, the seller determines the actual supply based on monopolistic profit maximization rules once bids are submitted. While this mechanism closely resembles the one used in China's SEO market, as outlined in Section \ref{SecTheory}, there are clear differences. In China's SEO format, the revenue to be raised is determined at the ex-ante stage, keeping the total supply fixed.

Building upon theoretical research, a series of empirical studies have been undertaken to examine the performance of uniform price auctions in the primary market for government bond issuance. \cite{umlauf1993empirical} analyzed Mexican government bond auction data from 1986 to 1991, comparing bid prices with secondary market prices to determine investor profits. Umlauf discovered that when Mexico's Treasury switched from discriminatory price auctions to uniform price auctions, investors made less profit, suggesting that uniform price auctions could generate more revenue for the issuer than discriminatory price auctions. However, \cite{nyborg1996discriminatory} did not find evidence in favor of uniform price auctions in the U.S. Treasury market, using when-issued or secondary market prices as benchmarks. This discrepancy could be due to the limited number of observations (only 15) of uniform price auctions in their sample. \cite{malvey1998uniform}, referencing when-issued market prices and using U.S. Treasury auction data, further explored uniform price auction performance in the U.S. market, finding evidence that issuer revenues are indeed higher under the uniform price mechanism. \cite{goldreich2007underpricing} analyzed U.S. Treasury bonds issued from June 1991 to December 2000, using the difference between issuance yield rates and when-issued market yield rates as a measure of discounting. The study found that bonds issued via the uniform price mechanism had a discount of 0.32 basis points, compared to 0.59 basis points for those issued via the discriminatory mechanism. This significant difference, even after controlling for various factors, provides strong empirical evidence that the uniform price mechanism can yield higher seller revenue.

However, utilizing the difference of primary market issuance prices relative to concurrent when-issued or secondary market prices as a proxy for seller revenue when evaluating uniform price auction performance has two potential issues: first, the when-issued or secondary market prices might be influenced by the primary market issuance prices; second, such analyses are only relevant for samples employing both the uniform price and discriminatory price auction mechanisms. To tackle these challenges, the academic community has pursued further investigations. \cite{keloharju2005strategic}, using Finnish government bond auction data, compared the actual investor bidding behavior to the theoretical predictions under the uniform price auction theory and found no evidence that rational investors were using market power to lower bid prices, contrary to what theories like those of \cite{back1993auctions} and \cite{wang2002auctioning} had anticipated. \cite{Kastl2011} employed a structural approach to deduce investors' valuation distribution from Czech government bond auction data, assuming that all investors' bid prices and quantities are in equilibrium. His analysis of the uniform price auction mechanism's operational efficiency suggested that it is highly effective in terms of maximizing issuer revenue and optimizing resource allocation efficiency, with sellers overpaying by no more than three basis points. Similarly, \cite{hortaccsu2018bid} analyzed U.S. Treasury auction data from July 2009 to October 2013, accounting for information asymmetry between primary dealers and other investors. They discovered that primary dealers, despite having higher valuations, submitted lower bids, resulting in an average investor surplus of just three basis points and a resource allocation efficiency loss of only two basis points. These results indicate that the uniform price auction mechanism is fundamentally effective in the U.S. Treasury market.

The exploration of the uniform price auction mechanism in the primary equity market dates back to the pioneering study by \cite{spatt1991preplay}, which first connected the IPO pricing process with uniform price auctions. They argue that to maximize issuance revenue, the IPO pricing process resembles a second-price sealed auction, particularly when underwriters set the offering price at the highest possible level below the market-clearing price following book building. 

Building on this foundation, subsequent studies have compared the efficacy of uniform price auctions against other stock issuance mechanisms. \cite{biais2002ipo} employed a unified analytical framework to demonstrate that standard uniform price auctions, where investors can act as monopolists within a specific price range, are less efficient due to potential implicit collusion and reduced pricing effectiveness compared to adjusted uniform price auctions and book building mechanisms. \cite{sherman2005global} argued that uniform price auctions are inherently less efficient because they do not allow underwriters to control the number of participating investors or the costs associated with information gathering. This perspective was further elaborated by \cite{jagannathan2015share}, who highlighted the intricacies faced by investors in standard uniform price auctions, such as the need to evaluate the company's value, understand competitor strategies, and devise bidding strategies, which collectively render the mechanism less efficient than book building in the IPO context.

However, \cite{kremer2004underpricing} countered that the step-function nature of the demand curve in standard uniform price auctions enables underwriters and issuers to reduce efficiency losses from investor market power by adjusting price and quantity intervals. Furthermore, critiques asserting the inefficiency of uniform price auctions often overlook potential agency conflicts between underwriters and issuers, where underwriters might act against the IPO company's interests. Addressing this, \cite{biais2002optimal} examined IPO pricing mechanisms amidst underwriter-issuer agency issues, deducing that adjusted uniform price auctions are optimal, particularly when accounting for potential collusion between underwriters and investors.

Empirical studies by \cite{derrien2003auctions} evaluated IPO discounts and volatilities across various mechanisms within the French market, identifying the lowest levels of both in IPOs utilizing adjusted uniform price auctions. \cite{kandel1999demand} explored the limit order books of 27 Israeli firms that conducted IPOs through standard uniform price auctions, observing a flat demand curve adjacent to the market-clearing price and identifying a direct relationship between initial return rates and demand price elasticity. Further, \cite{degeorge2010auctioned} investigated the behavior and outcomes of investors in 19 U.S. IPOs priced through adjusted uniform price auctions, highlighting issues such as overbidding by uninformed investors and significant demand price elasticity among institutional investors, which suggested the disclosure of proprietary information.

In summary, the literature assesses the efficacy of uniform price auctions in the primary market of financial securities, including bonds and equities, focusing on issuance revenue maximization, investor surpluses or returns, and information extraction. Theoretical studies typically explore how specific structural features of uniform price auctions influence auction outcomes, while empirical work examines the performance of uniform price auctions by comparing them with other issuance mechanisms or using a structural estimation methodology. Within studies on primary equity market issuance mechanisms, three main limitations emerge: Firstly, there's a noted scarcity of empirical studies on the performance of uniform price auctions in the primary equity market compared to bond markets. Secondly, the limited studies available rely on small samples, which may not provide a robust understanding of the auction mechanisms' performance and could potentially limit the generalizability of the findings. Lastly, there is a limitation in the theoretical literature regarding the impact of structural variations in uniform price auctions in practice on their efficacy, with existing theories suggesting that standard uniform price auctions might be less efficient than book building or adjusted uniform price auctions.

\section{Theoretical model}\label{SecTheory}

In this section we provide a preliminary model to study firm behavior is SEO auctions in China. Suppose there are $n>1$ firms (investors) in a financial market that are potentially interested in purchasing shares of a particular company called $S$. Potential buyers are indexed by $i\in \{1,...,n\}$. Firm $S$ issues a maximum of $\bar{Q}>0$ shares that are available for sell. Firm $S$ must raise an amount of money equal to $R>0$  by selling at most $\bar{Q}$ shares. Each share has a common value $v$ which is not known to any party at the time of offering. However, each representative firm $i$ receives a signal $s_i$ regarding the value $v$. Suppose all signals are independently and identically distributed according to some distribution function $F(.)$ with density $f<0$. Each bidder $i$'s marginal valuation for a share is then denoted by $v_i(s_i)$ which is a non decreasing and continuous function of the signal.

The seller employs a uniform-price auction with a specific variation to sell issued shares, where a maximum of $\bar{Q}$ shares will be available at a reserve price of $r$ and a target revenue $R$. After all bidders have submitted their bids, the auctioneer begins from the highest bid to clear the market. To achieve the target revenue, $R$, the seller follows this rule: starting from the highest bid, they calculate the revenue using the uniform pricing rule, whereby all bidders pay the same price, which is equal to the highest losing bid.\footnote{Should there be multiple bids at the clearing price resulting in excess demand, a \textit{pro rata} rationing rule will be applied.} The process continues to the next bid until either the revenue reaches $R$ or all available shares are sold. Furthermore, if the price reaches the reserve price, it will not decrease any further; the auction then concludes at a price equal to the reserve price, with the quantity sold being determined at the point where aggregate demand meets the reserve price. Here, ``aggregate demand'' refers to the total of all bids by all participants. To formally present the allocation and pricing rules, denote $q_i(p)$ as the demand schedule submitted by each bidder $i$, which is constructed to be non-increasing. Consequently, $Q(p) = \sum_{i \in I} q_i(p)$ represents the aggregate demand from all bidders. $R(p, Q)=p\times Q$ denotes the revenue at any given price $p$ and quantity $Q$. The auction clearing rule can thus be formally described as follows.

\begin{equation}\label{1}
      p^*  =  
      \begin{cases}
      & \max \{p \vert Q(p) < \bar{Q}\} \quad \text{if} \quad  R(p, Q(p)) = R \\
      & \max \{(p,r) \vert  Q(p) = \bar{Q} \: \text{or} \: Q(r) \leq \bar{Q} \} \quad \quad otherwise
    \end{cases}
\end{equation}

We would like to investigate both sellers and buyers strategic choices based on the above auction rules. First note that since for each bidder the marginal valuation for all the shares are equal, we expect the truthful bids for every firm to be a flat line. Of course, given firms have different signals regarding the values, the aggregate demand is not going to be flat, but rather a step function in which each step represents a truthful bid by a given firm. We formally define truthful bids as follows.

\begin{defn}
Truthful bids are denoted as flat bids by each firm at a price equal to their expected valuation $v_i(s_i)$.
\end{defn}

Once we define the truthfulness in the auction, we are going to investigate how China's variation of the SEO auction changes the bidding behavior of bidders. It is a well-known result in the common value auction literature that the uniform-price auction with fixed supply would result to untruthful bids and demand reduction \citep{back1993auctions, ausubel2014demand}. The variation in China's SEO, as described above, would set a maximum number of shares available, and would stop the allocation once the total revenue from the auction reaches $R$. 






\subsection{Seller's optimal strategy}
Based on the descriptions regarding China's SEO auctions, we are going to assume that the seller of shares (issuer) does not have any control over the choice of the reserve price. Also the funds required by the issuer or the target revenue $R$ is specified exogenously and is not a strategic choice of the issuer. However, the issue has control over the total number of shares issued for sale to achieve the target fund $R$. Therefore, based on the choice of the total shares available and the amount of money to raise we have either of the three possibilities below:

\begin{equation}
\begin{split}
   & (1) \: r\times \bar{Q} > R \\
     & (2) \: r\times \bar{Q} < R \\  
     &  (3) \: r\times \bar{Q} = R
\end{split}
\end{equation}

If the choice of the total number of shares is such that $r\times \bar{Q} > R$, then it means that the number of shares available is too large, and the actual number of shares to be sold would be definitely less than $\Bar{Q}$. This may seem like a counter-intuitive strategy since the price can never go below $r$. So the total number of shares would never be sold in the equilibrium.  In our data, 482 number of observations have this characteristic. One possible explanation of such a strategy is to signal that the seller is happy to sell all the shares at the reserve price.

If $r\times \bar{Q} < R$, then selling all the shares at the reserve price would fail to raise the fund amount required $R$. This essentially means that the seller is confident that the auction price is strictly above $r$. {\color{black}In our data, 197 number of auctions have this characteristic. This suggests that selling all shares at the reserve price would not fulfil the fundraising target ($R$). Therefore it indicates the seller's confidence that the auction price will exceed the reserve price.}

Finally, if $r\times \bar{Q} = R$, then the seller would either sells all the shares available of some amount less than that depending on the auction clearing price. This means that the seller is fine with the lower estimate of the value of shares being equal to $r$. {\color{black}The 475 SEOs are in the middle (considering 50,000,000 Yuan above or below equal to zero).  This suggests that the seller is content to either sell all available shares or a portion thereof, depending on the auction clearing price. }

We would like to first investigate the underlying reasons for the above behaviors and explore possible incentives that would result in such behaviors. First, suppose all buyers submit truthful bids. However, as suggested in the baseline model, since the signals are random draws buyers have different expected values and therefore submit different bids. The next proposition characterizes the best response of the seller in this situation.

\begin{pro}\label{pro1}
    When bids are submitted truthfully, the seller's optimal decision is to set $\bar{Q}$ such that, $r\times \bar{Q} = R$.
\end{pro}

\begin{proof}
Denote $b_i(p,q)$ as the bid submitted by bidder $i$ which specifies the bid price $p$ and the quantity demanded $q$. When bids are truthful $p$ is equal to the bidder's value, that is, $v(s_i)$. Our aim is to show the strategy $r\times \bar{Q} = R$ is at least as good as the other two. If all bidders bid truthfully then we denote $\mathbf{B}(s)$ as the vector of all bids submitted which represents the aggregate demand. Suppose the seller sets the maximum available quantity of shares $\bar{Q}$ such that $r\times \bar{Q} = R$. Given $\mathbf{B}(s)$ and $R$, if $p^*$ is such that $Q(p^*)\leq \bar{Q}$ and $p^* \times Q(p^*)=R$, then by setting a larger maximum number of shares, say $\bar{Q}_2>\bar{Q}$ where $r\times \bar{Q}_2 > R$, the seller's revenue would not change. However by setting a smaller maximum share $\bar{Q}_1<\bar{Q}$, such that $r\times \bar{Q}_1 < R$, the seller would receive a strictly lower revenue in cases where ${Q}_1 <Q(p^*)<\bar{Q}$. Also setting $r\times \bar{Q} > R$ is weakly dominated by $r\times \bar{Q} = R$ as the auction rules do not allow the price going below $r$.
\end{proof}

The above proposition suggests that if all bidders were to submit truthful bids then the seller's optimal strategy is to choose $\bar{Q}$ such that $r\times \bar{Q} = R$. The intuition behind this result is that when the seller knows bids are not strategic and truthful, then they would choose the total number of shares such that in the worst case scenario in terms of prices they would raise $R$. This is mainly because their main objective is to raise $R$. 

It is well known in the literature that it is a weakly dominant strategy for bidders to submit truthful bids for their first bid in a uniform-price auction \citep{Krishna2009}. However, this only holds for the first bid; for subsequent bids, bidders strictly prefer to bid lower than their true value. We refer to this as untruthful bidding. We would like to check how the new variations in the auction rule would change truthful bidding compared to the standard uniform-price auction. 

\subsection{Auction performance}

As explained before, $R$ is set before the bidding starts and the seller commits not to sell any more shares once $R$ is reached. We want to compare the bidding in this scenario to a case where there is no such target revenue and the seller simply makes $\bar{Q}$ units available in a uniform-price auction. We assume the seller, in case of commitment to a target revenue, sets $\bar{Q}$, such that $r\times \bar{Q} = R$. The next proposition shows in this case bidders bid more truthfully compared to a case where there is no target revenue. 

\begin{pro}\label{protruth}
    Committing to a target revenue would reduce the extent of untruthful bidding in the uniform-price auction. 
\end{pro}

\begin{proof}
 First denote $\mathbf{B}'(s)$ as an equilibrium of the uniform-price auction without a target revenue. We want to show when a target revenue $R$ is set, in the new equilibrium $\mathbf{B}(s)$, bids for all bidders are at least as large as the one in $\mathbf{B'}(s)$. Denote $\mathbf{b'_i}(s_i)$ as the vector of bids submitted by a representative bidder in $\mathbf{B}'(s)$. Given the reserve price $r$ there are two possibilities. First is that the equilibrium price of $\mathbf{B}'(s)$, say $p'$ is above the reserve price. To avoid trivial cases we assume the sum of demands by all bidders exceeds $\bar{Q}$. Therefore $\mathbf{B}'(s)$ results in all the $\bar{Q}$ units sold at $p'$.

 Now imagine a target revenue $r\times \bar{Q} = R$ is set. Given that $p'> r$ then it is straightforward to say $p'\times \bar{Q} > R$. Therefore there exist a $Q^*< \bar{Q}$ with a $p^*>p'$, such that $p^*\times Q^* = R$. Focusing on the representative bidder $i$, denote the new equilibrium bid as $\mathbf{b^*_i}(s_i)$. If $\mathbf{b^*_i} < \mathbf{b'_i}(s_i)$, given that $Q^*< \bar{Q}$ and $p^*>p'$, then bidder would definitely become worst off by submitting lower bids, as lower bids would reduce the probability of winning units without influencing the price paid. 

 The second possibility is when the equilibrium price of $\mathbf{B}'(s)$, $p'$ is equal to the reserve price. In this case the total units sold are either equal or less than $\bar{Q}$. When it is equal to $\bar{Q}$ setting a target revenue does not change the equilibrium as in the new equilibrium both price and quantity would become the same neither of the firms have incentives to deviate from $\mathbf{B}'(s)$. Also in case the quantity sold at the reserve price is below $\bar{Q}$, since the revenue target cannot be achieved, it is not binding.
\end{proof}

The above proposition suggests that the mechanism employed by China's SEO market, would reduce untruthful bidding. The intuition behind this result is as follows. When there is a target revenue, reducing bids below the expected valuation in the form of demand reduction comes at a cost: the cost of losing some units without influencing the auction clearing price.\footnote{The target revenue acts in a similar manner as ex-post revenue maximization where the seller defines the number of units sold after observing the bids using revenue maximization rules. See \cite{Back2001} and \cite{khezr2021revenue} for further discussions.} Therefore bidders would have less incentive to submit untruthful bids when target revenues are in place. 

Despite not specifying the extent of truthfulness, Proposition \ref{protruth} has a very important corollary. If bidders bid more truthfully in China's SEO variation compared to a standard uniform-price auction, it is possible that this variation results in higher revenues. In Section \ref{SecEmp}, we test this point empirically by providing a best-case scenario for the standard uniform-price auction and comparing it with the actual outcomes of China's SEO variation.

\subsection{Reserve price}

As explained in Section 2, the reserve price is determined by a specific rule, leaving firms with no control over it. The current rule in China's SEO market sets the reserve price at 90\% of the average share price in the 20 days preceding the pricing benchmark date. This raises the question of how auction outcomes might vary with changes to this rule. In particular, the existing rule aims to set the reserve price 10\% below the market's expected share price. Naturally, the reserve price cannot match the expected price, as this would exclude several bids including those with marginally lower expectations. The crucial issue is determining how much lower than the expected price the reserve price should be. There is no consensus in the literature regarding the optimal reserve price in uniform-price auctions, primarily due to the presence of multiple equilibria and a lack of understanding about optimal bidding behavior. A potential reference point could be the seller's valuation of the objects. However, in SEO markets, as in many others, sellers may prefer not to disclose their valuations, as this could influence buyers' bidding strategies.

Perhaps the rationale for establishing a specific rule regarding the reserve price is to standardize the benchmark across all auctions and mitigate the risk of significant losses or auction failures. It is still debatable whether setting the reserve price above or below 90\% would yield better outcomes. While a definitive answer to this debate is elusive, we can examine the advantages and disadvantages of adjusting the reserve price percentage. Setting the reserve price above 90\% of the average share price could increase the likelihood of excluding bids, potentially leading to more instances where the target revenue is not met. On the other hand, setting the reserve price below 90\% might increase the risk of reduced demand and lower share prices. This remains an interesting question for future research: how to compute the optimal reserve price given a revenue target in a uniform-price auction.

\section{Data}\label{SecData}
In this section, we undertake a rigorous process to ensure the reliability of our analysis. We describe the data collection process, variable definitions, and the cleaning procedures undertaken to ensure data quality and facilitate robust identification. Central to this section is the explication of our identification strategy. This strategy is key to establishing a clear cause-and-effect relationship between the variables. We will detail the methods employed to isolate the effect of interested variables on SEO discounts, minimizing the influence of confounding factors.
\subsection{Matching the SEO Data}
We consider all Chinese uniform-price auction SEOs from 8$^{th}$ May 2006 to 31$^{st}$December 2021. These are private placement SEOs.  To build effective metrics, we manually cross-check SEO information on the Oriental Fortune website and the WIND database, and the reports of SEO issuance on the Juchao Information website. Then, we collect investor bid price and location data for each SEO from the Non-Public Issuance Reports, Underwriter Compliance Reports, Law Firm Legal Opinions, and other documents from the Juchao Information website. In addition, Firm characteristics are from the WIND database. Thus we conduct our analysis at the bidder and SEO levels. So we deleted samples for the following reasons: (I) missing important variables value in both bidding and SEO level during the period; (II) not matching listed companies in the financial industry with (III) incomplete data. To this end, we use the following process:\\
(a) The data cleaning process prioritizes bidding data due to its importance in our analysis. This means we remove any SEO entries missing important bidding variables like reserve price, issue price, or shares to sell. Then to mitigate the influence of errors and outliers in the dataset, we apply Winsorization to the firm and deal with characteristics, setting limits at the 1st and 99th percentiles. Additionally, we exclude observations where the shares sold or to sell are not strictly positive. (b) To ensure accurate matching, exact SEO code “stkcd”, matches are confirmed between bidder and SEO levels. Then datasets are merged using a common "stkcd" code by iterating through each row of data, extracting bidding data and matching SEO data with the same SEO code and announcement date. Only rows with data and matching dates are retained, leading to the 36936 bidding samples from 1154 SEOs. (c) Finally, economic variables are incorporated from another dataset through a year-based merge.\\
Table \ref{ApProc} represents the sample selection process in each step of data preparation and matching data.\\
\begin{table}[h]
\centering
\caption{ The sample screening process is based on two levels of data.}
\label{ApProc}
\begin{tabular}{lc}
\hline
Panel A: Sample   screening process at the bidding level     & N                    \\ \hline                         
Investors’ bidding prices corresponding to SEOs using\\ auctions with bidding information in 2006-2021 & 47828                    \\
\multicolumn{2}{l}{Deleted:}                                                                               \\
Samples with missing important variables                                            & \multicolumn{1}{c}{5971} \\
SEOs with missing bidding information                                                               &  3336                    \\
\textbf{Final remaining samples}                                                    &            38521          \\ \hline
Panel B: Sample screening process at the   SEO level of non-public offering       & N                    \\ \hline
SEO samples using auctions with\\ bidding information in 2006-2021                                              &           1436           \\
\multicolumn{2}{l}{Deleted:}                                                                               \\
Samples with missing important variables                                            & \multicolumn{1}{c}{247} \\
Samples with outliers                                                               &           35           \\ 
\textbf{Final remaining samples}                                                    &             1154         \\ \hline
\end{tabular}

\end{table}
\subsection{Summary Statistics} 
Table \ref{Ap1} presents the summary statistics for various variables related to SEOs from 2006 to 2021. The "Offer discount" is the main variable of interest which has a mean value of 0.14 with a standard deviation of 0.10, indicating that, on average, the discount offered during an SEO is approximately 14\% relative to the closing price, with a range from -0.28 to 0.87. This indicates that some offerings were made below the market price (negative discount), while others were made above it. The "Issue price" has a mean value of 19.29 with a standard deviation of 30.65, suggesting considerable variability in the issue prices. This shows that the price at which shares are offered to the public can vary significantly, reflecting differences in market conditions, company valuation, and other factors. Similarly, the "Money raised" variable has a mean value of 14.56 and a standard deviation of 21, indicating the average amount raised during an SEO is around 14.56 units, but there is a wide dispersion in the amounts raised. Further, the average number of shares to sell is approximately 18,174, while the average number of shares sold is around 12,614. This suggests that, on average, companies tend to sell a portion of the shares they initially plan to offer, possibly due to market demand or other strategic considerations.\\
\begin{table}[H]
\centering
\caption{Descriptive statistics for SEO from 2007 to 2021. The definition of the above variables is presented in Table \ref{VarDef}.}
\label{Ap1} 
\begin{tabular}{lcccccccc}
\hline
Variable           & n    & mean  & sd    & median & min   & max    & skew  & kurtosis \\
\hline
Reserve price      & 1154 & 16.88 & 26.41 & 11.77  & 2.27  & 550.24 & 12.85 & 226.62 \\
Issue price        & 1154 & 19.29 & 30.65 & 13.22  & 2.41  & 645.00 & 13.16 & 235.54 \\
Money to raise     & 1154 & 15.96 & 22.61 & 9.06   & 0.36  & 220.00 & 4.50  & 26.54 \\
Money raised       & 1154 & 14.56 & 21.00 & 8.09   & 0.32  & 200.00 & 4.52  & 26.86 \\
Shares to sell     & 1154 & 18174.19 & 27953.46 & 9000.00 & 264.82 & 250000.00 & 4.01 & 21.06 \\
Shares sold        & 1154 & 12614.85 & 20656.36 & 6128.76 & 165.45 & 247934.00 & 4.52 & 28.87 \\
Offer discount     & 1154 & 0.14 & 0.10 & 0.14      & -0.28 & 0.87  & 0.17  & 2.96 \\
Presd30            & 1154 & 0.03 & 0.01 & 0.03      & 0.01  & 0.07  & 0.98  & 0.74 \\
Close price        & 1154 & 22.32 & 34.83 & 15.48   & 2.67  & 711.21 & 12.60 & 218.99 \\
Offer size         & 1154 & 0.17 & 0.26 & 0.13      & 0.00  & 5.31  & 14.25 & 265.91 \\
Market value       & 1154 & 22.89 & 1.01 & 22.73    & 20.48 & 26.64 & 0.60   & 0.20 \\
Underwriter        & 1154 & 0.45  & 0.50 & 0.00     & 0.00  & 1.00  & 0.18  & -1.97 \\
Turnover20         & 1154 & 0.25  & 0.46 & 0.04     & 0.00  & 3.41  & 3.21  & 13.15 \\
Precar5            & 1154 & 0.02  & 0.06 & 0.01     & -0.23 & 0.41  & 0.95  & 3.68 \\
Analyst            & 1154 & 1.80  & 1.15 & 1.95     & 0.00  & 4.14  & -0.19 & -1.08 \\
Institution        & 1154 & 13.75 & 24.16 & 0.16    & 0.00  & 92.44 & 1.57  & 1.07 \\
Migration          & 1154 & 0.05  & 0.12 & 0.00     & 0.00  & 1.00  & 3.12  & 11.22 \\
Intratio           & 1154 & 0.09  & 0.12 & 0.04     & 0.00  & 1.00  & 1.88  & 5.29 \\
Investornum        & 1154 & 2.28  & 1.10 & 2.40     & 0.00  & 4.60  & -0.45 & -0.34 \\
Leverage           & 1154 & 0.48  & 0.20 & 0.48     & 0.06  & 1.70  & 0.30  & 0.46 \\
ROA                & 1154 & 0.05  & 0.06 & 0.04     & -0.76 & 0.48  & -1.65 & 35.08 \\
GDP growth         & 1154 & 0.10  & 0.04 & 0.09     & 0.02  & 0.22  & 0.33  & -0.10 \\
Market size        & 1154 & 0.66  & 0.14 & 0.68     & 0.38  & 1.21  & -0.16 & 0.54 \\
\hline
\end{tabular}

\end{table}
To account for heterogeneity in the relationship between discount and other factors, we incorporate interaction terms into our regression models. These terms capture how the effect of one variable on discount can be influenced by another variable. For example, the interaction term "Money to raise" with "Shares to sell" (MtoR) allows us to explore how the relationship between offering a discount and the amount of money needed changes based on the number of shares being sold. Similarly, the "Return on Assets" with "Analyst Coverage" (RoAn) interaction term examines how the impact of a firm's profitability (ROA) on discounts varies depending on whether analysts cover the firm. This could indicate that the effect of ROA is stronger (or weaker) for companies with more analyst attention. Finally, the "Pre-money Valuation" with "Leverage" (PrLe) interaction term investigates how the influence of a firm's pre-money valuation on discounts is contingent on its leverage ratio. This helps us understand how a company's financial structure moderates the effect of certain characteristics. By including these interaction terms, we go beyond examining individual variables and capture the complex and nuanced relationships that may exist within the data. We will discuss further about these terms further in the main regression models in subsection \ref{regsec}.

\subsection{Identification and Variables selection}
In our analysis, we employ a technique known as identification analysis to assess whether the relationships between variables within the random forest model can be uniquely isolated. This step of analysis involves identifying the variables that are relevant to the research question and ensuring that the model is well specified to capture the relationships of interest. This involves determining if a change observed in one variable can be definitively attributed to a change in another, and if this cause-and-effect association can be estimated without any ambiguity.\footnote{ In line with econometric principles, our identification analysis investigates whether the random forest model allows for the estimation of causal effects. This translates to pinpointing if changes in the independent variables definitively lead to changes in the dependent variables.} This step is essential to avoid overfitting. Overfitting can occur when a model is too complex, including variables that do not add meaningful explanatory power to the model. \\
By leveraging the Random forest's ability to explore multiple decision trees with randomized feature selection, we aim to achieve robust identification and minimize the influence of confounding factors that might otherwise obscure the true underlying relationships.\footnote{Random forest is a systematic approach that explores diverse variable combinations, treating each as a distinct model with unique sets of data points. For further details, refer to \citep{james2013introduction}.}\\
There are several advantages of using the random forest technique instead of other statistical methods like Stepwise Regression and Regularization Techniques (Lasso/Ridge regression). It can capture complex relationships (even non-linear) and interactions between variables effectively. Further, it provides a measure of feature importance, ranking variables based on their contribution to the dependent variable in the model. 

\begin{figure}[h]
\centering
\begin{tabular}{@{}cc@{}}
\includegraphics[width=0.5\textwidth, height=7.5cm]{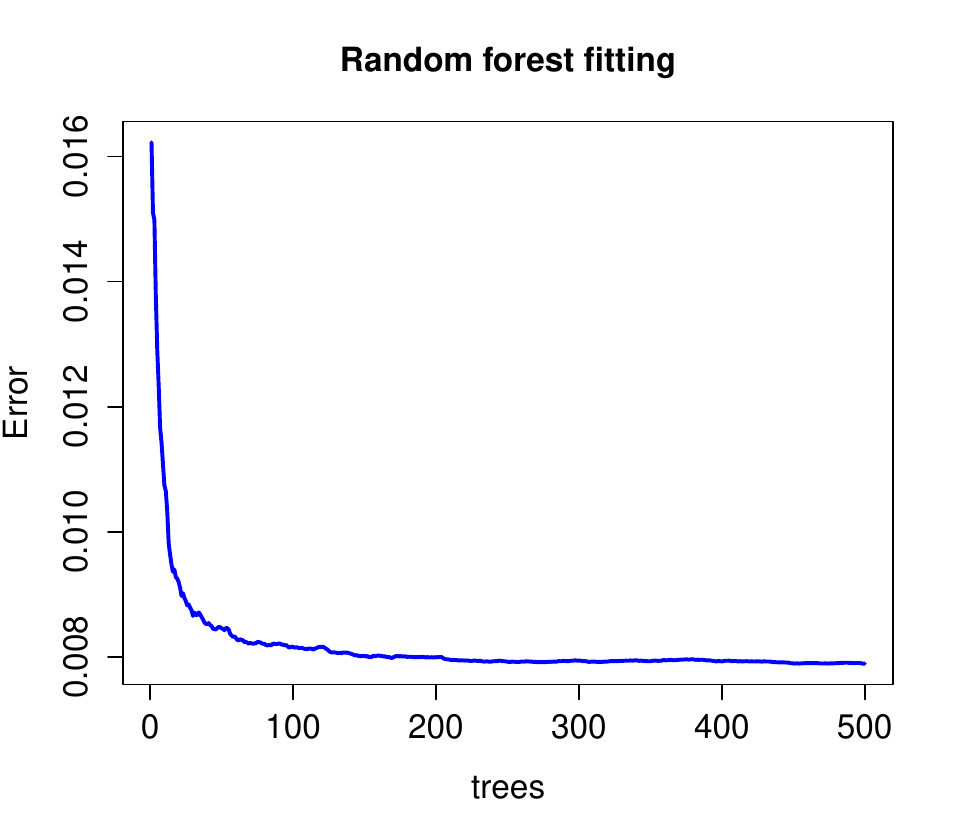} &
 \includegraphics[width=0.5\textwidth, height=7.1cm]{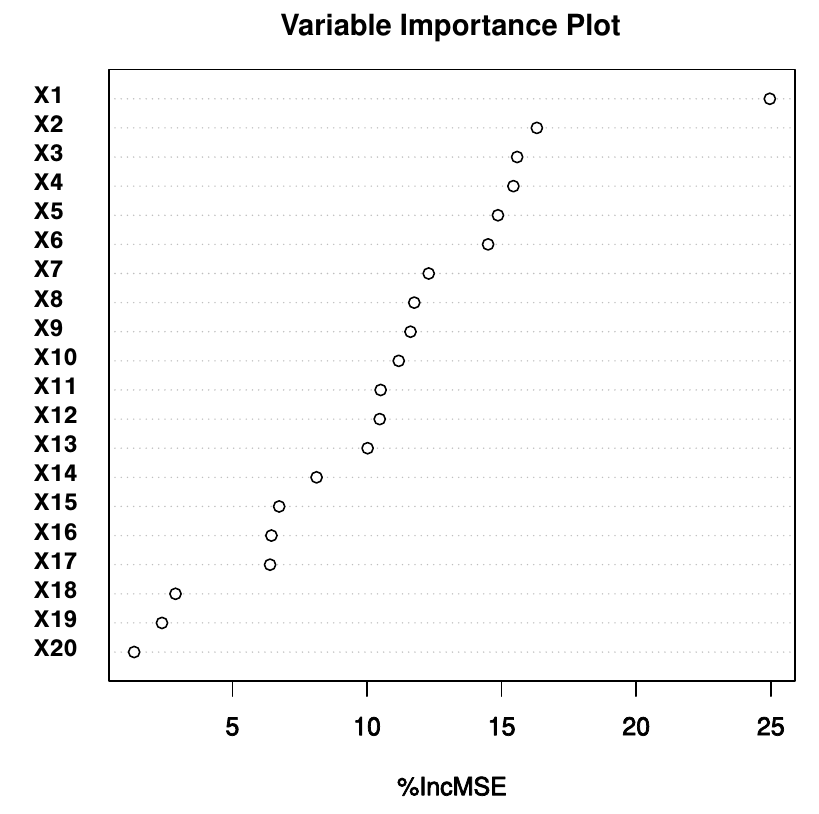}\\  
   \end{tabular}
\caption{Plots from parameter tuning in Random Forest algorithm determine the optimal number of trees and variables. }
\label{RF}
 \end{figure}

The Random Forest algorithm results, displayed in Figure \ref{RF}, reveal insights into various variables. The left plot demonstrates the mean squared residuals stabilizing around 0.008 after approximately 250 iterations. The right plot depicts variable weights, highlighting their proportional impact on Mean Squared Error (MSE). Notably,  Institution, Share to sell, and ROA. The first 13 variables were deemed significant for predicting the outcome, while the latter variables did not contribute substantially to the predictive power, (variables from $X_{14}$ (Year) to $X_{20}$ (Migration)). This prioritization highlights the importance of the first 13 variables in explaining the variance in the response variable, warranting their inclusion in the model. Note that we attempted to include additional variables in the model; however, many of them did not reach statistical significance and did not improve the adjusted $R^2$. To mitigate the risk of overfitting, we decided to retain only the first 13 variables in the main model. It is worth mentioning that, unlike simple regression models, Random Forest regression models do not provide coefficients; instead, they consist of a collection of decision trees, each constructed using a random subset of predictors. Consequently, Random Forest regression models estimate a set of weights representing the significance of each predictor. This approach mitigates the risk of overfitting and enhances sensitivity to outliers.\footnote{This approach helps avoid overfitting because the model is not overly reliant on any single variable. It is also less sensitive to outliers that might skew coefficients in traditional models.}
\section{Empirical Modelling and Analysis}\label{SecEmp}

 \setlength{\tabcolsep}{7.5pt}
 For precision and analysis of bidder behavior, we categorised the dataset into three groups based on the relationship between `money to raise' (R), the `reserve price' (r), the average number of shares available (Q) and $\bar{Q}$ `shares to sell'. The first group (482 observations) represents situations where the number of shares offered (r x Q) significantly exceeds the required funding ($r\times \bar{Q}- (10,000 \times R)>1000$). The second group (197 observations) reflects scenarios where selling all shares at the reserve price wouldn't meet the funding target ($r\times \bar{Q}- (10,000 \times R)< -1000$). Finally, the third group (475 observations) captures situations where the difference between offered shares and required funding falls within a \$1000 threshold ($-1000<r\times \bar{Q}- (10,000 \times R)<1000$), suggesting bidders are comfortable with the reserve price as a reasonable valuation.\footnote{We chose the threshold of \$1000 based on the scaling of the variables involved in our analysis. The total number of shares is represented by the product of the quantity and 10,000 shares ($(\bar{Q}*10000)$). The issue price is in Yuan, and the money to raise is denoted in 100 million Chinese Yuan (R $\times$ 100,000,000). Considering these scales and data formats, our threshold equation $(r*\bar{Q}*10,000-R*100,000,000>10,000,000)$ which simplifies to $(r*\bar{Q}-R*10000>1000)$. We also tested alternative thresholds, such as 5000 or 500, and found consistent and non-significant changes in the results, affirming the robustness of our chosen threshold.}\\
Further, to investigate the behavior of bidders, we defined two dummy variables; if the expression $(\text{{reserve price}} \times \text{{shares to sell}} - (\text{{money to raise}} \times 10,000)) > 1000$, then `Exceed reserve` is set to 1; otherwise, it is set to 0. Similarly, if the expression $(\text{{reserve price}} \times \text{{shares to sell}} - (\text{{money to raise}} \times 10,000)) < -1000$, then `Below reserve` is set to 1; otherwise, it is set to 0. These dummy variables aim to capture specific bidding behaviors related to the reserve price, shares to sell, and the amount of money to raise, helping to analyze and understand the bidder strategies and preferences in the auction scenarios.
\subsection{Regression models}\label{regsec}

The following multiple regression model aims to investigate the result of Proposition \ref{pro1}.
\begin{equation}\label{mainequ}
    Discount_{i,t}=\beta_{0}+\beta_{1}\textit{Exceed reserve}_{i,t}+\beta_{2}\textit{Below reserve}_{i,t}+\beta_{i} \sum controls_{i,t}+ \sum Ind+ \epsilon_{i,t},
\end{equation}
where SEO discount (Discount), is $100*(CLS – OFFER)/CLS$; where OFFER represents the SEO offer price and CLS denotes the market stock price one day before the SEO. Consistent with the literature on SEO pricing \citep{corwin2003determinants, huang2011managing, sun2023sunshine}, controls represent a set of control variables such as the cumulative abnormal return of the stock before the SEO (CAR),  analyst following (Analyst), offer size (Offersize), institutional investor ownership (Institution), reserve price, money to raise plus three interaction terms, Termt, Terms and Termc, help to capture the complex and nuanced relationships that may exist within the data, revealing insights that may not be apparent when considering the variables individually or in isolation.

We present the regression results for Equation \ref{mainequ} in Table \ref{T1}. In column (1), we observe results from a simplified regression focusing solely on the explanatory variable, while column (2) presents a more comprehensive regression incorporating control variables. In investigating the impact of various bidding auction policies on firm-level SEO discount (Discount), our attention is drawn to the variables ``Exceed reserve" and ``Below reserve". In both cases, their coefficients are statistically significant and positive, indicating meaningful effects on the SEO discount. Specifically, "Exceed reserve" has a coefficient of 0.143, and "Below reserve" has a coefficient of 0.130. Moving to column (2), similar patterns emerge, with both variables maintaining statistically significant coefficients of 0.028 and 0.030, respectively. These results suggest that deviating from the optimal strategy specified in Proposition \ref{pro1} comes at the cost of lower share prices on average. 

Additionally, the variable ``Price Difference Ratio" exhibits statistical significance with a coefficient of 0.026, as anticipated. Among the control variables, coefficients for Institution, Offersize, money to raise, Investornum, Anayst, and ROA are positive and significant, while those for TERMT are TERMS are negative and statistically significant.

Furthermore, we assessed multicollinearity within the model by calculating the variance inflation factor (VIF).\footnote{The VIF measures how much the variance of the estimated regression coefficient is increased due to collinearity among the predictor variables.} The average VIF value, computed at 3.86, suggests a lack of significant collinearity between the control variables and the variables of interest.
\subsection{Robustness checks}
In our robustness analysis, we conduct two key tests to ensure the reliability and consistency of our findings regarding SEO discounts. Firstly, we explore the impact of alternative bidding investors by separating our analysis based on bidder type, examining bids from both individual and institutional investors. Secondly, we investigate the influence of different time horizons on SEO discounts, considering periods before and during the COVID pandemic. Through these tests, we aim to provide a comprehensive understanding of the factors affecting SEO discounts, addressing potential variations in bidding behavior and market conditions. 

\setlength{\tabcolsep}{36pt}
\begin{table}[H]
    \centering
    \caption{ This table presents the ordinary least squares results for the impact of different policies of bidding auctions on firm 
level SEO discount (Discount). Appendix \ref{Ap1} presents the variable definitions. Standard errors are in parentheses; and Signif. codes are  0 ‘***’ 0.001 ‘**’ 0.01 ‘*’ 0.05 ‘.’ 0.1, respectively.}
\label{T1}

\begin{tabular}{lcc}
\hline
Variable           & Model 0                                                    & Model 1                                                       \\ \hline
Exceed reserve                 &   \begin{tabular}[c]{@{}c@{}}0.143***\\ (0.006)\end{tabular}                                                         & \begin{tabular}[c]{@{}c@{}}0.028***\\ (0.007)\end{tabular}    \\
Below reserve                 & \begin{tabular}[c]{@{}c@{}}0.132***\\ (0.010)\end{tabular}                                                           & \begin{tabular}[c]{@{}c@{}}0.030***\\ (0.009)\end{tabular}    \\
money to raise     &  & \begin{tabular}[c]{@{}c@{}}0.000*\\ (0.000)\end{tabular}      \\
Price Difference Ratio                &  & \begin{tabular}[c]{@{}c@{}}0.026**\\ (0.009)\end{tabular}     \\
shares to sell     &                                                            & \begin{tabular}[c]{@{}c@{}}0.000***\\ (0.000)\end{tabular}    \\
Offersize          &                                                            & \begin{tabular}[c]{@{}c@{}}0.069***\\ (0.012)\end{tabular}    \\
PRECAR5            &                                                            & \begin{tabular}[c]{@{}c@{}}0.232$^{.}$\\ (0.139)\end{tabular} \\
Analyst            &                                                            & \begin{tabular}[c]{@{}c@{}}0.007*\\ (0.003)\end{tabular}      \\
Institution        &                                                            & \begin{tabular}[c]{@{}c@{}}0.001***\\ (0.000)\end{tabular}    \\
Investornum        &                                                            & \begin{tabular}[c]{@{}c@{}}0.023***\\ (0.002)\end{tabular}    \\
ROA                &                                                            & \begin{tabular}[c]{@{}c@{}}0.259**\\ (0.079)\end{tabular}     \\
TERMT              &                                                            & \begin{tabular}[c]{@{}c@{}}-0.000***\\ (0.000)\end{tabular}   \\
TERMS              &                                                            & \begin{tabular}[c]{@{}c@{}}-0.168***\\ (0.039)\end{tabular}   \\
TERMC              &                                                            & \begin{tabular}[c]{@{}c@{}}-0.131\\ (0.252)\end{tabular}      \\ \hline
p-value            &   $<$ 2.2e-16                                                         & $<$ 2.2e-16                                                               \\
Adjusted R-squared & 0.407                                                      & 0.613        \\ \hline                                                
\end{tabular}
\end{table}

\subsubsection{Alternative investors for SEO discounts}
As a first robustness test, we construct SEO discounts conditional on alternative bidding investors. Our data includes bids from both individual and institutional investors. Assuming an efficient market, bid prices should reflect all available information regardless of the bidder type. Consequently, both individual and institutional investors are likely to make bidding decisions based on the same information.
To explore potential differences in bidding behavior, we separate the analysis based on bidder type. We estimate Equation \ref{mainequ} for two samples: bids with only institutional investors and bids with a mix of individual and institutional investors. This approach allows us to investigate whether the coefficients for exceeding or falling below the reserve price differ between these two bidder groups. The outcomes are detailed in Table \ref{RobT}.

\setlength{\tabcolsep}{14pt}
\begin{table}[H]
    \centering
    \caption{ This table reports the results of a robustness check for the impact of different policies of bidding auctions on firm 
level SEO discount (Discount),  based on different 
subsamples. Standard errors are in parentheses; and Signif. codes are  0 ‘***’ 0.001 ‘**’ 0.01 ‘*’ 0.05 ‘.’ 0.1, respectively.}
\label{RobT}
\begin{tabular}{lcccc}
\hline
                   & \multicolumn{2}{c}{Model 0}                                                                                                        & \multicolumn{2}{c}{Model 1}                                                                                                        \\ \hline
Variable           & \begin{tabular}[c]{@{}c@{}}Institutional \\ investor\end{tabular} & \begin{tabular}[c]{@{}c@{}}Individual \\ investor\end{tabular} & \begin{tabular}[c]{@{}c@{}}Institutional \\ investor\end{tabular} & \begin{tabular}[c]{@{}c@{}}Individual \\ investor\end{tabular} \\ \hline
Exceed reserve                 & \begin{tabular}[c]{@{}c@{}}0.143***\\ (0.006)\end{tabular}        & \begin{tabular}[c]{@{}c@{}}0.142***\\ (0.006)\end{tabular}     & \begin{tabular}[c]{@{}c@{}}0.028***\\ (0.007)\end{tabular}        & \begin{tabular}[c]{@{}c@{}}0.025***\\ (0.007)\end{tabular}     \\
Below reserve                 & \begin{tabular}[c]{@{}c@{}}0.131***\\ (0.010)\end{tabular}        & \begin{tabular}[c]{@{}c@{}}0.130***\\ (0.010)\end{tabular}     & \begin{tabular}[c]{@{}c@{}}0.028**\\ (0.009)\end{tabular}         & \begin{tabular}[c]{@{}c@{}}0.028**\\ (0.010)\end{tabular}      \\
money to raise     &                                                                   &                                                                & \begin{tabular}[c]{@{}c@{}}0.000*\\ (0.000)\end{tabular}          & \begin{tabular}[c]{@{}c@{}}0.000$^{.}$\\ (0.000)\end{tabular}  \\
Price Difference Ratio                &                                                                   &                                                                & \begin{tabular}[c]{@{}c@{}}0.026**\\ (0.009)\end{tabular}         & \begin{tabular}[c]{@{}c@{}}0.022*\\ (0.009)\end{tabular}       \\
shares to sell     &                                                                   &                                                                & \begin{tabular}[c]{@{}c@{}}0.000***\\ (0.000)\end{tabular}        & \begin{tabular}[c]{@{}c@{}}0.000***\\ (0.000)\end{tabular}     \\

OFFERSIZE          &                                                                   &                                                                & \begin{tabular}[c]{@{}c@{}}0.065***\\ (0.012)\end{tabular}        & \begin{tabular}[c]{@{}c@{}}0.138***\\ (0.021)\end{tabular}     \\
PRECAR5            &                                                                   &                                                                & \begin{tabular}[c]{@{}c@{}}0.237$^{.}$\\ (0.140)\end{tabular}     & \begin{tabular}[c]{@{}c@{}}0.350*\\ (0.150)\end{tabular}       \\
Analyst            &                                                                   &                                                                & \begin{tabular}[c]{@{}c@{}}0.008*\\ (0.003)\end{tabular}          & \begin{tabular}[c]{@{}c@{}}0.004\\ (0.003)\end{tabular}        \\
Institution        &                                                                   &                                                                & \begin{tabular}[c]{@{}c@{}}0.001***\\ (0.000)\end{tabular}        & \begin{tabular}[c]{@{}c@{}}0.001***\\ (0.000)\end{tabular}     \\
Investornum        &                                                                   &                                                                & \begin{tabular}[c]{@{}c@{}}0.022***\\ (0.003)\end{tabular}        & \begin{tabular}[c]{@{}c@{}}0.022***\\ (0.003)\end{tabular}     \\
ROA                &                                                                   &                                                                & \begin{tabular}[c]{@{}c@{}}0.284***\\ (0.081)\end{tabular}        & \begin{tabular}[c]{@{}c@{}}0.183*\\ (0.072)\end{tabular}       \\
TERMT              &                                                                   &                                                                & \begin{tabular}[c]{@{}c@{}}-0.000***\\ (0.000)\end{tabular}       & \begin{tabular}[c]{@{}c@{}}-0.000**\\ (0.000)\end{tabular}     \\
TERMS              &                                                                   &                                                                & \begin{tabular}[c]{@{}c@{}}-0.176***\\ (0.041)\end{tabular}       & \begin{tabular}[c]{@{}c@{}}-0.095*\\ (0.037)\end{tabular}      \\
TERMC              &                                                                   &                                                                & \begin{tabular}[c]{@{}c@{}}-0.066\\ (0.254)\end{tabular}          & \begin{tabular}[c]{@{}c@{}}-0.542$^{.}$\\ (0.283)\end{tabular} \\ \hline
p-value            &   $<$ 2.2e-16                                                                 &  $<$ 2.2e-16                                                               &     $<$ 2.2e-16                                                               &  $<$ 2.2e-16                                                               \\
Adjusted R-squared & 0.3905                                                            & 0.399                                                          & 0.6072                                                            & 0.6265    \\ \hline                                                    
\end{tabular}

\end{table}

The findings regarding "Exceed reserve" and "Below reserve" exhibit consistency across both institutional and individual investor subsamples, as observed in the coefficients reported. These coefficients, approximately 0.143 and 0.131 in the solo model without control variables, and 0.028 in the main model with additional control variables, remain statistically significant. This suggests a notable impact of these deviations on the SEO discount, irrespective of investor type, aligning with the results observed in the full sample as depicted in Table \ref{T1}.

Of particular interest is the similarity in the coefficients for "Exceed reserve" and "Below reserve" between both investor types. The similarity, with coefficients of approximately 0.143 and 0.131 in the solo model and 0.028 in the main model, underscores the notion that both institutional and individual investors react similarly to deviations from the reserve price when making bidding decisions in SEO auctions. Additionally, the consistency in the levels of statistical significance among control variables in both scenarios further supports this observation, with values being closely aligned across the two investor types.

Overall, the results suggest that while there might be a minor difference in how institutional and individual investors react to the reserve price, their bidding behavior is largely similar. This aligns with the assumption of an efficient market where all available information, including the reserve price, is reflected in bid prices.

\subsubsection{Alternative horizons for SEO discounts}
\begin{figure}[H]
\centering
   \begin{tabular}{@{}cc@{}}
     \includegraphics[width=0.485\textwidth, height=5.13cm]{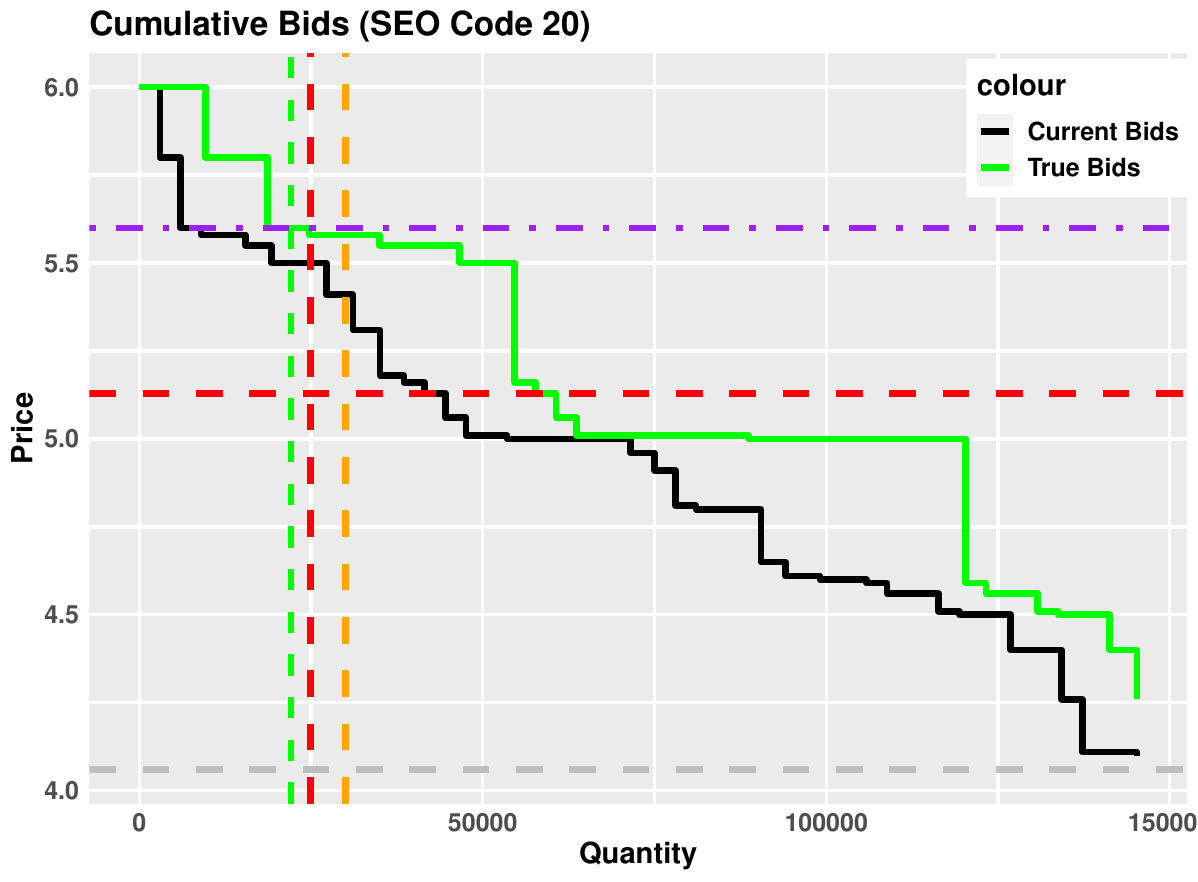}&
  \includegraphics[width=0.485\textwidth, height=5.13cm]{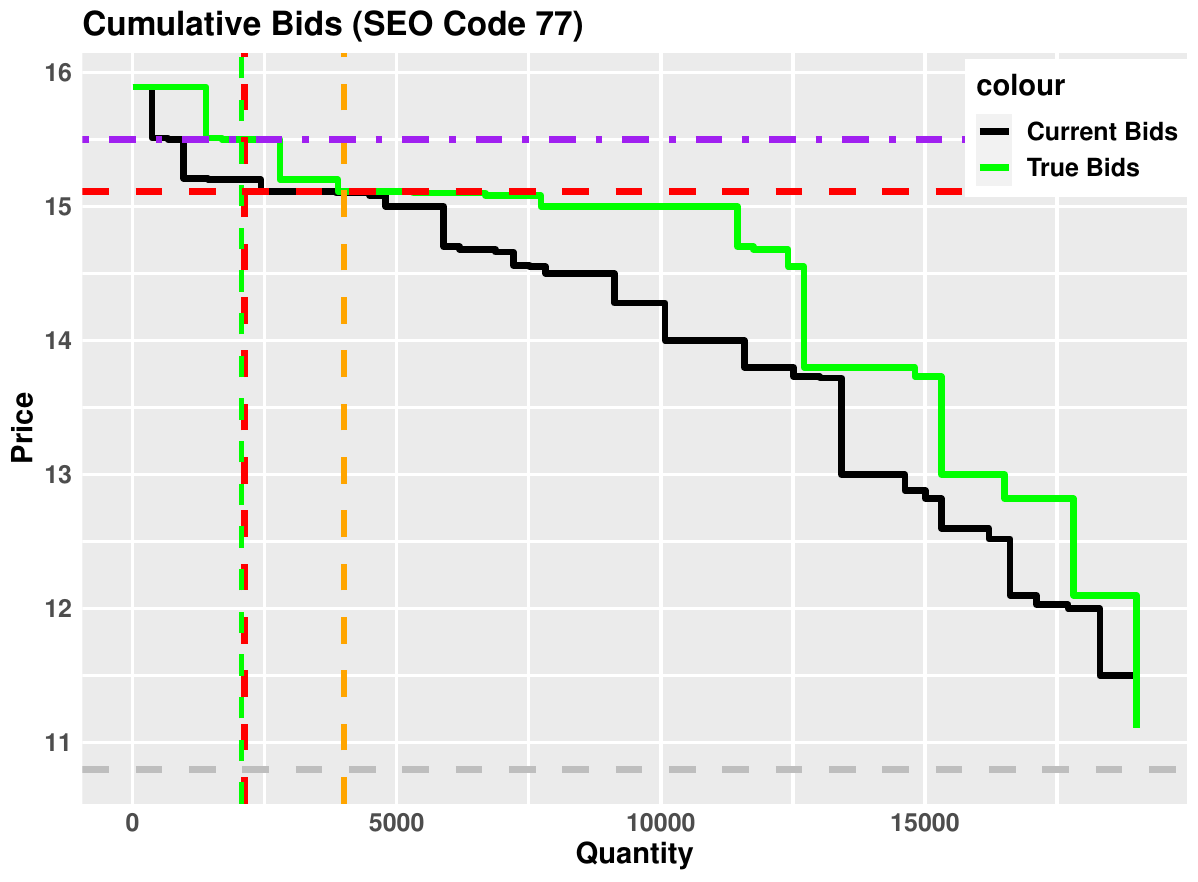} \\
     \includegraphics[width=0.485\textwidth, height=5.13cm]{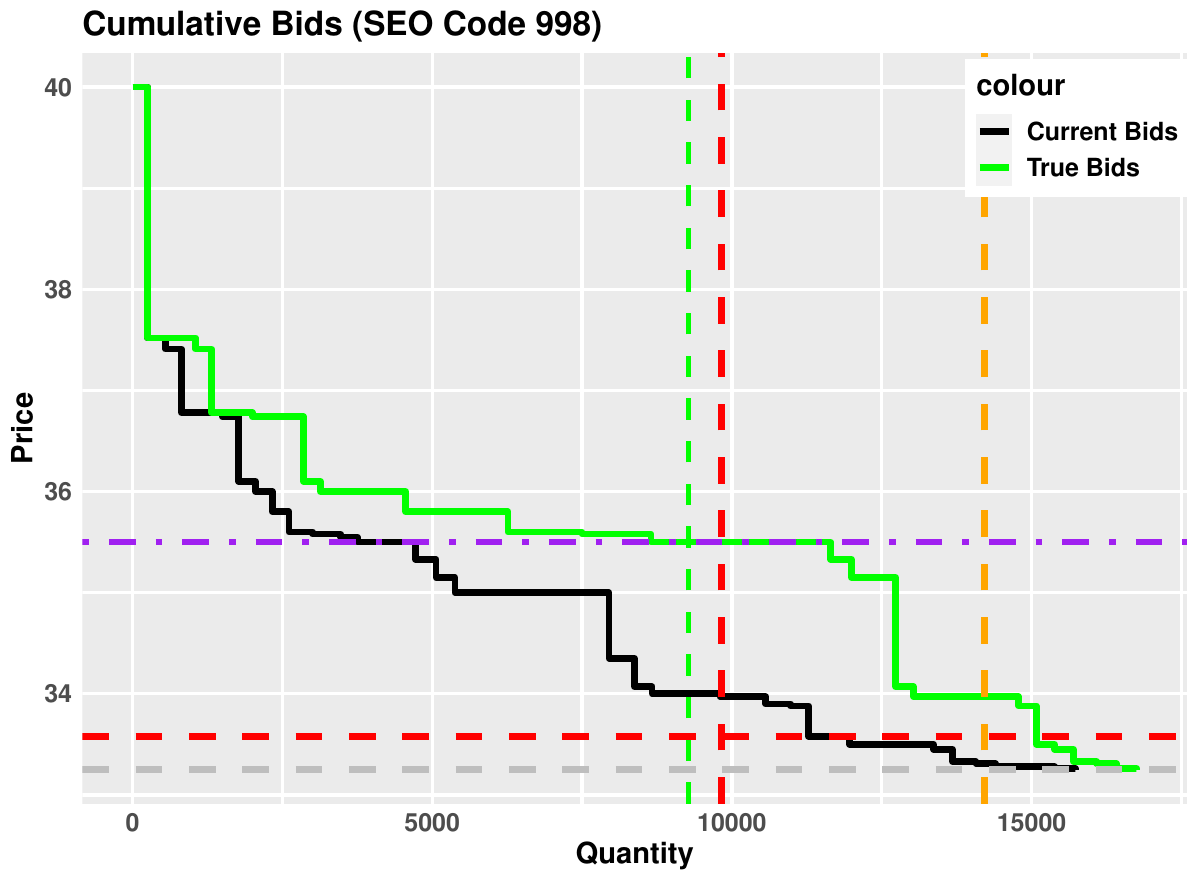}&
  \includegraphics[width=0.485\textwidth, height=5.13cm]{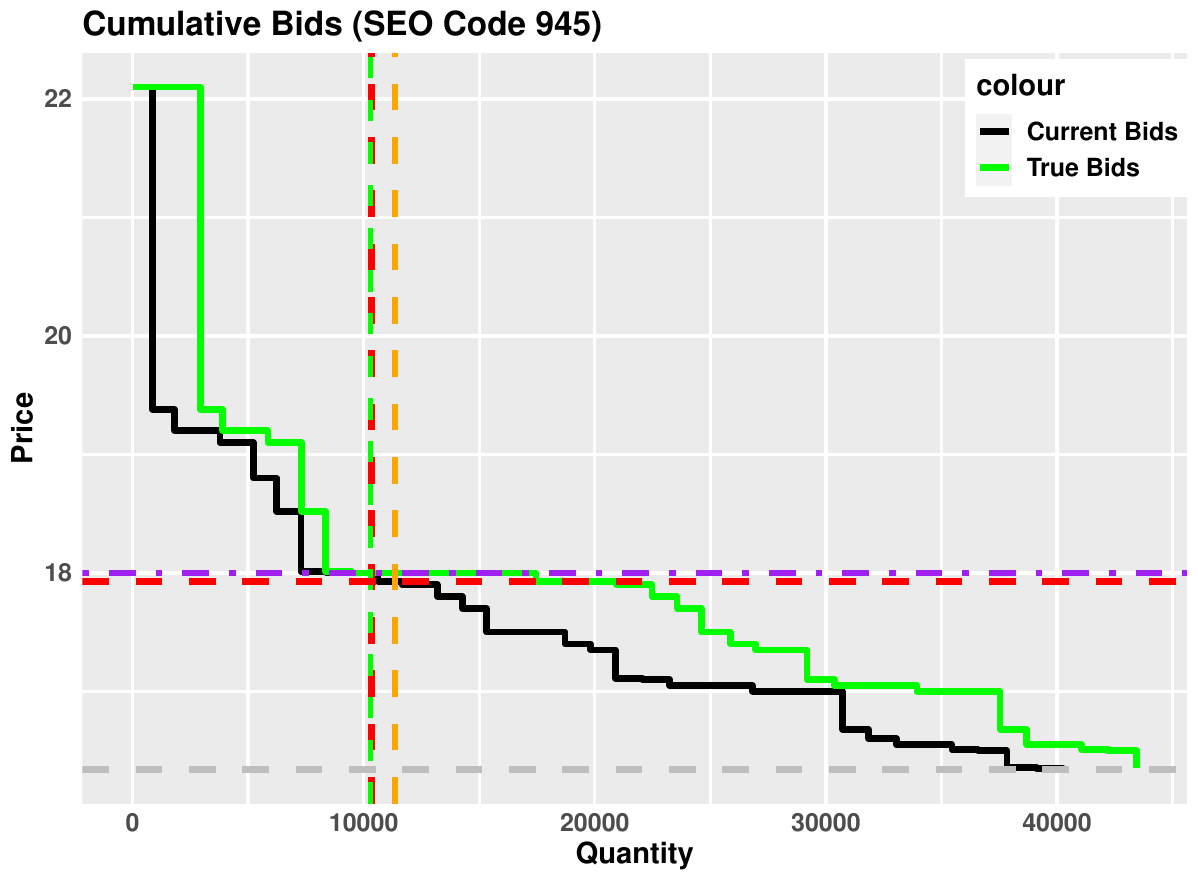} \\
   \includegraphics[width=0.485\textwidth, height=5.13cm]{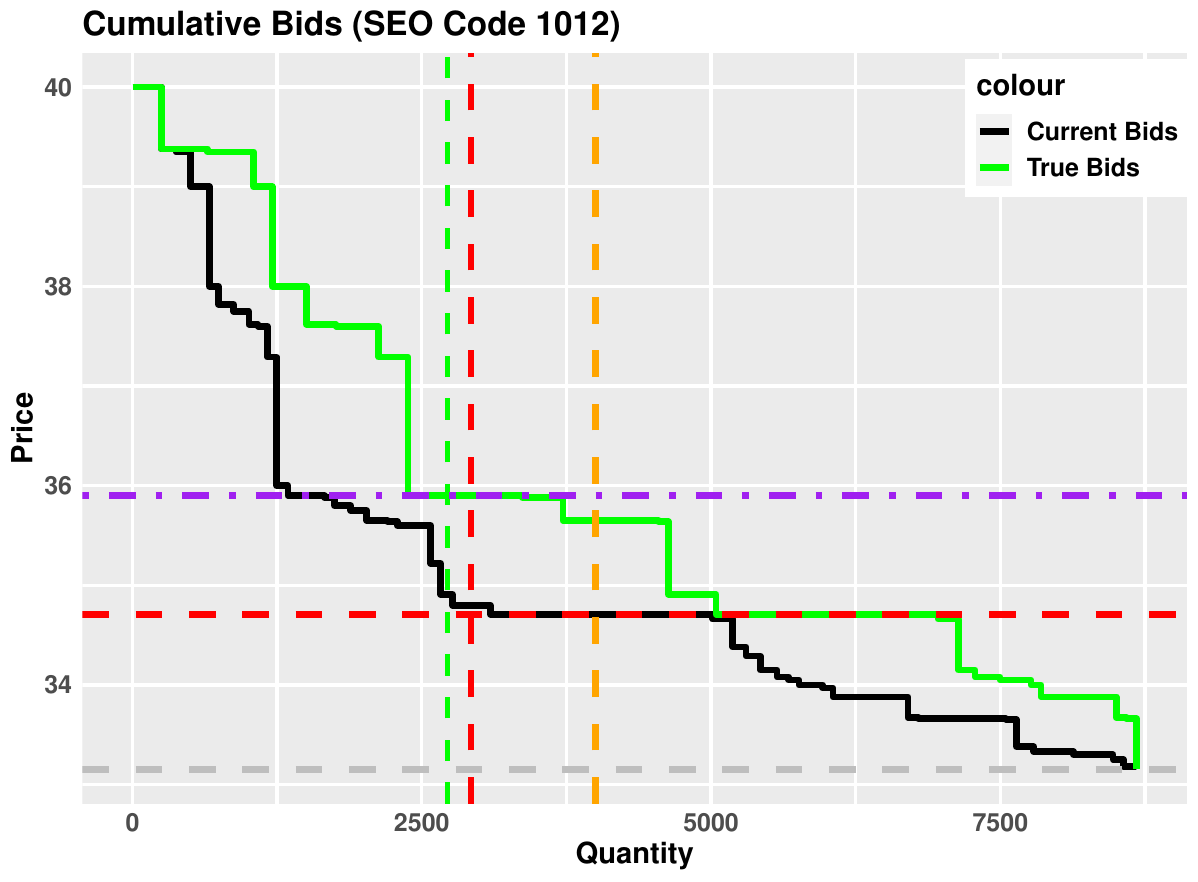} & 
  \includegraphics[width=0.485\textwidth, height=5.13cm]{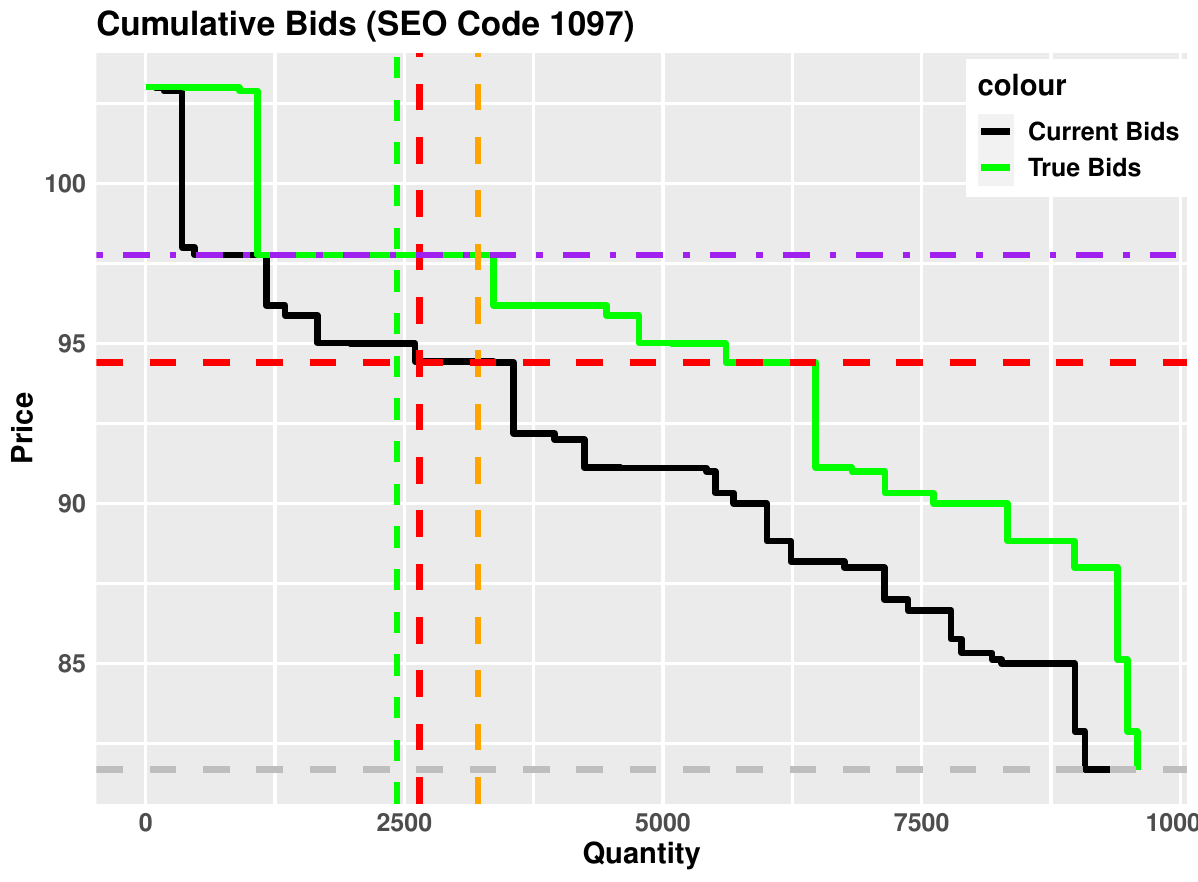}  
   \end{tabular}
\caption{Cumulative demand across various SEOs is displayed. The vertical red dashed line indicates the shares already sold, while the vertical orange dashed line shows the shares still available for sale. The vertical green dashed line represents the hypothetical demand if bidders bid truthfully. Additionally, the horizontal purple dot-dash line also depicts a hypothetical price in truthful bidding. The horizontal grey dashed line signifies the reserve price.} 
\label{figTruth}
 \end{figure}

As a second robustness test, we construct SEO discounts conditional on alternative horizons.  We divided the data into two subsamples: one spanning from February 2007 to December 2019, pre-dating the COVID-19 pandemic, comprising 820 SEOs; and the other from January 2020 to December 2021, covering the period during and after the pandemic, containing 334 SEOs. Models (1) and (3) represent Equation \ref{mainequ} without control variables, while models (2) and (4) include control variables. Notably, the coefficients for ``Exceed reserve" and ``Below reserve" remain consistently positive and statistically significant across all four columns. This consistency supports the finding of Proposition \ref{pro1}.  \\
In this section, we empirically investigate the bidding behavior and the performance of the auction mechanism in China's SEO market. Given that our current data includes bids and auction outcomes for every auction, we first construct a hypothetical case to forecast truthful bids (aggregate demand). As highlighted before, it is well-known that in standard uniform-price auctions, bidders often submit bids that are lower than their actual values \citep{Krishna2009}. However, in both the standard and adjusted variations, it remains a weakly dominant strategy for bidders to submit their true value as their first bid. Using this insight, we assume each player's first bid is their true expected value for the shares. We then construct the truthful aggregate demand for each auction based on these first bids. For example, as shown in Figure \ref{figTruth}, the green curve represents truthful bids based on a few representative samples. This curve is obviously at least as high as the submitted demand shown in black, as the first bids are at least as high as the subsequent bids for each bidder.

\setlength{\tabcolsep}{18pt}
\begin{table}[h]
    \centering
    \caption{ This table reports the results of a robustness check for the impact of different policies of bidding auctions on firm 
level SEO discount (Discount),  based on different 
subsamples. Standard errors are in parentheses; and Signif. codes are  0 ‘***’ 0.001 ‘**’ 0.01 ‘*’ 0.05 ‘.’ 0.1, respectively.}
\label{RobT2}
\begin{tabular}{ccccc}
\hline
            & \multicolumn{2}{c}{2007-2019}                                                                                          & \multicolumn{2}{c}{2020-2021}                                                                                           \\ \hline
Variable    & (1)                                                          & (2)                                                         & (3)                                                          & (4)                                                          \\ \hline
Exceed reserve          & \begin{tabular}[c]{@{}c@{}}0.130***\\ (0.009)\end{tabular} & \begin{tabular}[c]{@{}c@{}}0.022**\\ (0.008)\end{tabular} & \begin{tabular}[c]{@{}c@{}}0.157***\\ (0.007)\end{tabular} & \begin{tabular}[c]{@{}c@{}}0.066***\\ (0.011)\end{tabular} \\
Below reserve         & \begin{tabular}[c]{@{}c@{}}0.130***\\ (0.011)\end{tabular} & \begin{tabular}[c]{@{}c@{}}0.024*\\ (0.010)\end{tabular}  & \begin{tabular}[c]{@{}c@{}}0.138***\\ (0.016)\end{tabular} & \begin{tabular}[c]{@{}c@{}}0.043**\\ (0.016)\end{tabular}  \\
Control     & NO                                                         & YES                                                       & NO                                                         & YES                                                        \\
Industry    & YES                                                        & YES                                                       & YES                                                        & YES                                                        \\
N           & 820                                                        & 820                                                       & 334                                                        & 334                                                        \\ \hline
p-value & $<$ 2.2e-16                                                      & $<$ 2.2e-16                                                     & $<$ 2.2e-16                                                      & $<$ 2.2e-16     \\
Adjusted R-squared & 0.2962                                                     & 0.5823                                                    & 0.6119                                                     & 0.7657    \\ \hline                                                
\end{tabular}

\end{table}

The truthful bids construct a benchmark for the analysis in this section. We use this benchmark to first investigate what the auction price would be if the bidders were to submit truthful bids. Then, we use this benchmark to construct the truthful equilibrium outcome with the total number of shares available, creating a metric for a scenario where a standard uniform-price auction without the revenue target variation is conducted.

\setlength{\tabcolsep}{9pt}
 
\subsection{The effect of bids on auction price}

Having created the hypothetical aggregate demand based on the truthful bids, we then examine what the auction clearing price would be if the auctioneer were to raise the same amount of target revenue but with truthful bids instead. The auction clearing price and the quantity sold based on this hypothetical scenario are represented by the horizontal purple and vertical green dashed lines in Figure \ref{figTruth}.

Our empirical analysis shows that the difference between the price in the case of truthful bidding and the real issue price is less than \$1 (specifically, \$0.88). Additionally, the price difference ratio is 0.028, which confirms that the revenue raised by the actual auction is very close to a truthful bid scenario. This gives further support to the result of Proposition 2. While we were unable to provide an actual extent for the truthful bids in that proposition, our empirical results suggests the China's SEO variation of the uniform price auction is almost truthful. 

\subsection{The performance of the auction mechanism}

To analyze the performance of the auction adjustment, we propose a hypothetical scenario where all available shares are sold in a standard uniform-price auction with a set reserve price. In such an auction, there is no revenue target, and therefore the auction clearing price is determined by the intersection of the demand schedule with supply (depicted as a vertical line at $\bar{Q}$). If this hypothetical scenario were implemented, the demand schedule submitted by bidders would likely change. To address this, we create a 'best-case scenario' and compare it to the current mechanism.

\begin{figure}[h]
\centering
   \begin{tabular}{@{}cc@{}}
     \includegraphics[width=0.495\textwidth, height=6.3cm]{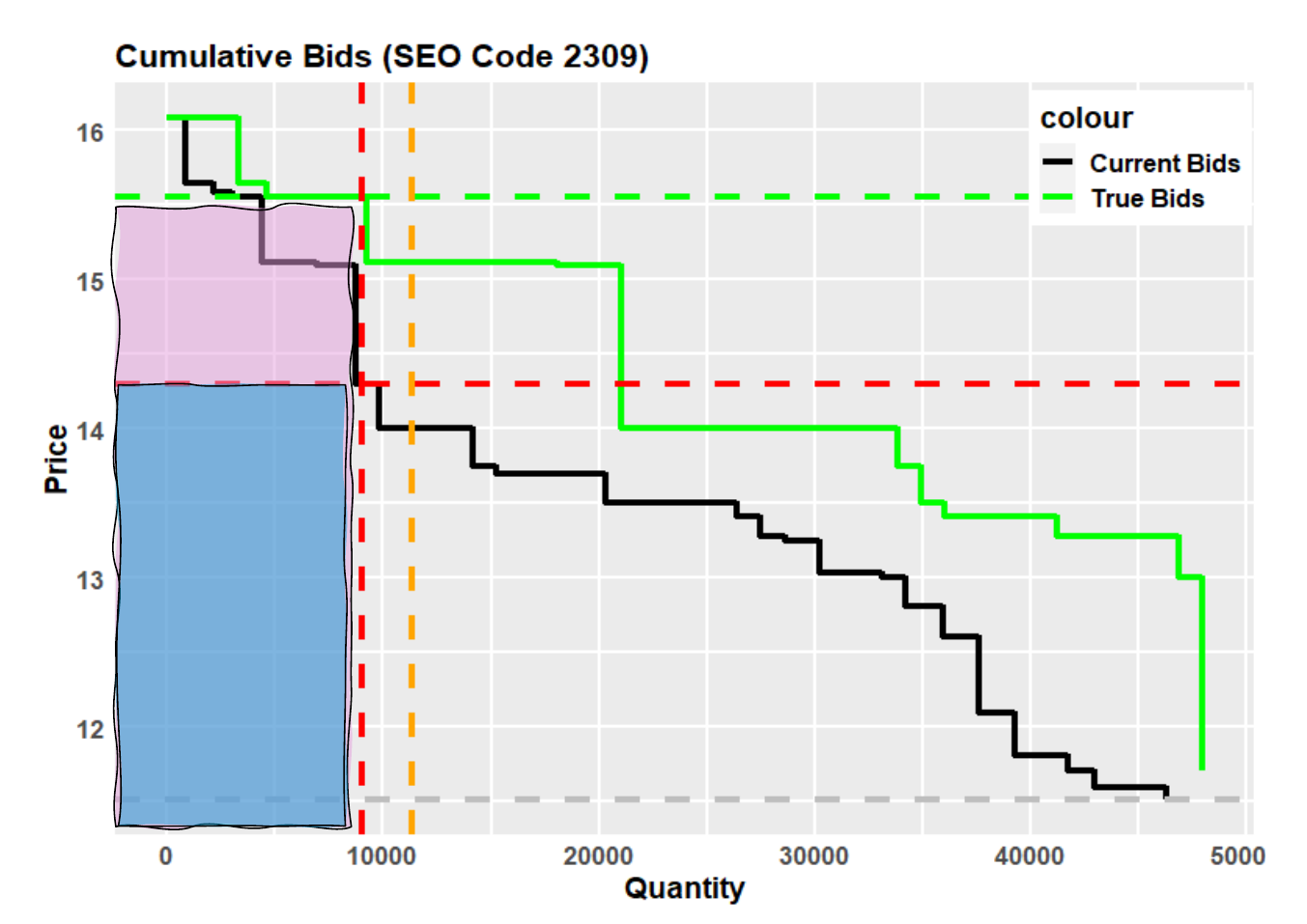}&
  \includegraphics[width=0.495\textwidth, height=6.3cm]{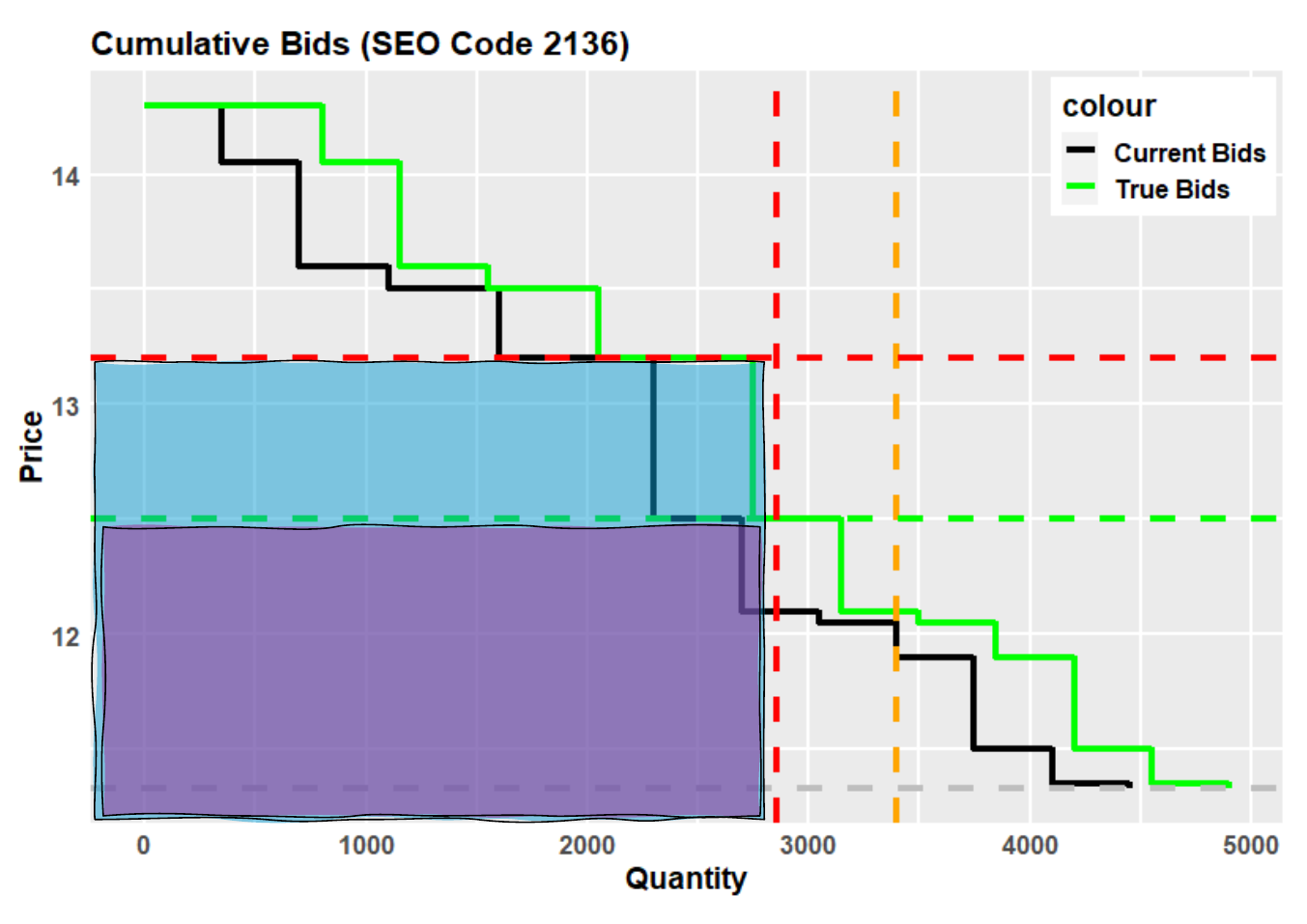} \\  
   \end{tabular}
\caption{Cumulative demand is depicted across different SEOs. The vertical red dashed line denotes shares sold, while the vertical orange dashed line indicates shares to sell. The green horizontal dot-dash line represents the hypothetical line. Additionally, the horizontal red dashed line symbolizes the issue price.} 
\label{figAreas}
 \end{figure}

Using the truthful bids, we construct the best-case scenario for the standard uniform-price auction with $\bar{Q}$ units. Assuming the first bid from each bidder is their true expected value, we establish a truthful demand schedule for all participants (see Figure \ref{figAreas}). It is important to note that in the actual scenario, the number of shares sold is less than or equal to $\bar{Q}$, while in the hypothetical scenario, the total quantity is always $\bar{Q}$.

To address this issue, we assume that for any additional unit that an issuer does not sell in the auction, they could obtain the same surplus (revenue) as in the best-case scenario, that is, the truthful equilibrium price. This means both actual and hypothetical scenarios receive the same revenues for all unsold units, $\bar{Q}-Q$. Therefore, when we compare the revenue of the two cases, we only compute the revenue for the actual sold units. Specifically, for the actual scenario observed in the data, the revenue is equal to the number of shares sold times the auction clearing price (issue price). For the hypothetical scenario, we first compute the truthful equilibrium price using the truthful demand and $\bar{Q}$. Then we compute the revenue by multiplying the truthful price by the number of shares sold in the actual auction, that is, $Q$.


Figure \ref{figAreas}, shows cumulative demand curves for two scenarios: one based on real bids and the other on truthful bids. In the truthfulness scenario, represented by the horizontal green dot-dash line, we see the potential price range. Meanwhile, the horizontal red dashed line indicates the price set for SEO release. Additionally, the red vertical dashed line represents the current quantity sold, and the orange vertical dashed line signifies the shares available for sale. Finally, the horizontal grey dashed line denotes the reserve price.  

To compare the revenue from both scenarios, we employ cubic spline interpolation to create smooth curves for the cumulative bids (for both black and green cumulative curves in Figure \ref{figAreas}). The Pink Area is determined by computing the cumulative area under the truth cumulative curve (depicted in green) and the vertical line corresponding to the maximum number of shares to sell. This vertical line, known as the 'hypothetical line' in this context, is represented by the dark blue dot-dash line. The Pink Area calculation employs the trapezoidal rule, where adjacent points on the interpolated curve and the hypothetical line form trapezoids, (in some cases trapezoids will be converted to rectangles like the left plot in Figure \ref{figAreas}). 
The Actual Area encompasses the cumulative area under the actual bid curve up to the maximum number of shares sold, multiplied by the issue price. Additionally, it includes the area under the true bid curve and the two vertical lines representing shares to sell and shares sold values.\\

\setlength{\tabcolsep}{9pt}
To assess the ratio of the difference between the forecasting (hypothetical area) and actual area, we employ three distinct criteria: Mean Absolute Percentage Error (MAPE), Mean Percentage Error (MPE) and Mean Forecast Bias (MFB).\footnote{Note that other criteria such as Mean Absolute Error are omitted due to their lack of ratio-based interpretation, potentially resulting in large values without specific meaning. Nevertheless, results for these metrics are available upon request.} MAPE quantifies errors as a percentage of the actual values, improving interpretability and mitigating sensitivity to data scale. It is defined as follows:
$$ MAPE = \frac{1}{n} \sum_{i=1}^{n} \frac{|y_i - \hat{y}_i|}{|y_i|} \times 100\%,$$
where  $y_i$ is the actual value and $\hat{y}_i$ is the forecasting value (in this case hypothetical value), MPE calculates the average percentage difference between the forecasted and actual values. Formula:
$$ MPE = \frac{1}{n} \sum_{i=1}^{n} \frac{y_i - \hat{y}_i}{y_i} \times 100\%. $$
The mean Forecast Bias (MFB) reveals both the direction and magnitude of bias in your forecasts relative to the actual values, calculated as follows:
$$MFB = \frac{1}{n} \sum_{i=1}^{n} (y_i - \hat{y}_i).$$ 
Our findings indicate that MAPE and MPE average approximately 2.91 \% and  -0.78\%, respectively. Additionally, the negative MFB, with a value of -1447.183, indicates that the actual values are, on average, 3\% below the hypothetical values.
The above result suggests that China's SEO variation of the uniform-price performed very-well in terms of revenue generation. This indicates that the revenue generated in the auctions is almost the same as what they could have obtained if bidders were all bidding truthfully.\\


\section{Conclusion}\label{ConSec}
In this study, we have explored the effectiveness of a variation of the uniform-price auction in China's seasoned equity offerings (SEO) market, where issuers commit to a pre-announced revenue target. Using a common value framework, we analyzed the operation, allocation of shares, and pricing of this auction mechanism. Our theoretical findings suggest that when buyers bid truthfully, the optimal strategy for the seller is to set the total share quantity such that it equals the target revenue divided by the reserve price. This approach appears to encourage a higher level of truthful bidding compared to a standard uniform-price auction without revenue commitment.

Empirical analysis using data from China's SEO market confirms our theoretical predictions. Our investigation showed that the variation of the uniform-price auction with a revenue target performs well in terms of revenue generation, with results comparable to those that would be achieved if all bidders submitted truthful bids. In particular, by constructing hypothetical scenarios and comparing them to actual auction outcomes, we demonstrated that China's SEO auctions generate revenue effectively and align closely with the theoretical best-case scenario.

The contributions of this paper are threefold. First, we advance the theoretical literature on uniform-price auctions by examining the implications of incorporating a revenue target. Second, we provide empirical evidence on the performance of SEO auctions in China, a market that offers a unique institutional setting for such analysis. Our study fills a gap in the literature by focusing on the primary equity market, where uniform-price auctions have been less commonly studied compared to bond markets. Third, we present evidence that, contrary to prior theoretical concerns, uniform-price auctions can perform well in primary equity markets. This insight challenges the conventional wisdom that the complexity of uniform-price auctions hinders their effectiveness for new share sales.

In conclusion, our findings highlight the potential benefits of using a uniform-price auction mechanism with a revenue target in the SEO market. This approach not only enhances revenue generation but also fosters a more truthful bidding environment. Future research could further investigate the optimal reserve price given a revenue target and explore additional variations of auction mechanisms to improve market efficiency and issuer outcomes.


\newpage
\bibliographystyle{econ}
\bibliography{SEO}
\newpage
\begin{appendices}


\renewcommand{\thefigure}{A\arabic{figure}}

\setcounter{figure}{0}

\renewcommand{\thetable}{A\arabic{table}}

\setcounter{table}{0}

\renewcommand{\theequation}{A\arabic{equation}}

\setcounter{equation}{0}

\section{Variable Definitions} \label{AppML}
\begin{table}[H]
\centering
\caption{Variable Definitions. Note: This table defines all variables used in the baseline regression.}
\label{VarDef}
\begin{tabular}{lp{12cm}}
\hline
\textbf{Variable} & \textbf{Definition} \\
\hline
Analyst \& Logarithm of 1 plus the number of analysts following the SEO firm. \\

Below reserve \& Indicates whether the funds raised fall short of the expected proceeds from share sales at the reserve price by more than a threshold. \\

Exceed reserve \& Indicate whether the funds raised exceed the expected proceeds from share sales at the reserve price by more than a threshold. \\

Institution \& Institutional ownership. \\

Intdum \& Dummy variable taking the value of 1 if the bid price is round, and 0 otherwise. \\

Intratio \& The number of round bids over total bids. \\

Investornum \& Logarithm of the total number of participants in an SEO. \\

Issue price \& The price at which a company offers its shares to investors during an SEO. \\

Lnprice1 \& Logarithm of the closing price one day before an SEO. \\

Migration \& The subscription ratio of controlling shareholders in an SEO. \\

Market value \& Logarithm of the market value. \\

Money to raise \& The total amount of funds a company aims to raise through an SEO. The scale of Money to raise is 100 million Yuan in Chinese. \\

Offerdiscount1 \& $\frac{(CLOSE1 - OFFERPRICE)}{CLOSE1}$. \\

Offersize \& Offer size of an SEO over total outstanding shares. \\

Precar5 \& Cumulative abnormal returns over 5 trading days before an SEO. \\

Presd30 \& Standard deviation of stock returns over 30 trading days before an SEO. \\

Price difference ratio \& The ratio of the difference between the issue price and the reserve price to the reserve price in a financial context. \\

Reserve price \& The minimum price below which shares will not be sold to investors during the offering. \\

Shares to sell & The total number of shares that a company intends to offer for sale during an SEO. The scale of Shares to sell is in millions (Million shares). \\

Turnover20 & Average turnover over 20 trading days before an SEO. \\

Underwriter & Dummy variable equal to 1 if the underwriter is ranked in the top 10 based on revenue, and 0 otherwise. \\
\hline
\end{tabular}

\end{table}


\end{appendices}

\end{document}